\documentclass[]{openjournal}

\usepackage{latexsym}
\usepackage{graphicx}
\usepackage{amssymb}
\usepackage{longtable}
\usepackage{amsmath}
\usepackage{graphicx}
\usepackage{hyperref}
\usepackage{newtxtext,newtxmath}

\slugcomment{Accepted for publication in The Open Journal of Astrophysics}
\shorttitle{Merger fractions in AGN host galaxies}
\shortauthors{Villforth}

\begin{document}

\title{A complete catalogue of merger fractions in AGN hosts: No evidence for an increase in detected merger fraction with AGN luminosity}
\author{C. Villforth$^{1}$ \thanks{E-mail: c.villforth@bath.ac.uk}}
\affil{$^1$University of Bath, Department of Physics, Claverton Down, BA2 7AY, United Kingdom \textit{c.villforth@bath.ac.uk}}

\maketitle

Despite the importance of Active Galactic Nuclei (AGN) in galaxy evolution, the mechanisms that fuel AGN activity remain poorly understood. Theoretical models suggest that major mergers of galaxies contribute strongly to AGN fuelling, particularly at high AGN luminosities. The connection between mergers and AGN activity has therefore been widely studied, although with contradictory results. Some studies find a strong connection between mergers and AGN, while others find merger fractions in AGN hosts to match those in the inactive galaxy population. To address these apparent contradictions, I present a complete and systematic analysis of detected merger fractions in AGN hosts from the literature. I assess if discrepancies between studies are indicative of systematic uncertainties and biases and analyse the detected merger fraction as a function of luminosity, redshift, and AGN selection method. X-ray selected AGN samples show comparable detected merger fractions across studies and major mergers do not dominate triggering in this AGN population. On the other hand, signatures of significant merger contribution to the AGN population are observed in a small fraction of primarily radio selected and reddened AGN samples. It is unclear if this is due to observational biases or physical differences in the host galaxies. There is no correlation between the detected merger fraction and AGN luminosity. This lack of correlation between detected merger fraction and AGN luminosity, which has previously been reported in the literature, cannot be explained by systematic uncertainties and observational biases. 

\vspace{1cm}

\twocolumngrid

\section{Introduction}
\label{S:intro}

Supermassive black holes are found in the centres of practically all massive galaxies \citep{kormendy_coevolution_2013}. The majority of supermassive black holes are quiescent, only a small fraction of black holes are observed to be actively accreting gas. These objects are known as Active Galactic Nuclei (AGN). Observations have suggested a close link between the growth of supermassive black holes and the galaxies they reside in. Throughout the history of the Universe, the black hole accretion rate density closely traces the star formation rate density \citep[e.g.][]{madau_cosmic_2014, aird_evolution_2010} and at low redshift, the black hole mass is correlated with the properties of the host galaxy \citep[e.g.][]{kormendy_coevolution_2013,gebhardt_relationship_2000,tremaine_slope_2002,novak_correlations_2006}. These data suggest a physical co-evolution between supermassive black holes and their host galaxies, although a purely statistical co-evolution is also consistent with observations \citep{jahnke_non-causal_2011}.

A popular theoretical model for black hole galaxy co-evolution was suggested initially by \citet{sanders_ultraluminous_1988}. In this model, black holes and galaxies experience significant growth during major mergers of galaxies. The merger initiates a central starburst, gas is then further funneled toward the central black hole. Accretion onto the black hole starts during an initially obscured phase. As the black hole accretion rate rises, surrounding gas is expelled through AGN feedback, revealing an unobscured AGN. As the feedback clears the surrounding gas, star formation is shut down. This co-evolution model has been popular in the literature and is supported by simulations \citep[e.g.][and references therein]{hopkins_characteristic_2009,di_matteo_energy_2005,somerville_semi-analytic_2008,alexander_what_2012}.

Studies of host galaxies of AGN have therefore often focussed on identifying a possible link between black hole growth and major galaxy mergers \citep[e.g.][]{bahcall_hubble_1997,canalizo_quasi-stellar_2001,veilleux_deep_2009,kocevski_candels:_2012,
schawinski_role_2010,schawinski_heavily_2012,grogin_agn_2005,
villforth_morphologies_2014,ellison_galaxy_2011,
ellison_galaxy_2013,ellison_galaxy_2015,ellison_definitive_2019,
koss_merging_2010,villforth_host_2017,villforth_host_2019,
mechtley_most_2016,treister_major_2012,glikman_major_2015,
urrutia_evidence_2008,boehm_agn_2012}.
Early studies of host galaxies of local luminous AGN with ongoing starbursts found high incidences of merger features \citep{canalizo_quasi-stellar_2001}. \citet{veilleux_spitzer_2009} showed that in the local population of ultra-luminous infrared galaxies (ULIRGs) and AGN, mergers are prevalent in sources showing strong starbursts, but merger fractions are low in the AGN population as a whole. Later studies found low fractions of mergers in AGN hosts \citep{dunlop_quasars_2003}, although some studies showed that merger features became prevalent in deeper imaging \citep{bennert_evidence_2008}. Further studies targeted a large number of moderate luminosity AGN in deep fields \citep[e.g.][]{georgakakis_host_2009,kocevski_candels:_2012, villforth_morphologies_2014,hewlett_redshift_2017} or targeted high luminosity or other rare AGN \citep{urrutia_evidence_2008, chiaberge_radio_2015, villforth_host_2017, mechtley_most_2016, marian_major_2019, villforth_host_2019}, comparing detected merger fractions in AGN hosts to those in control samples of inactive galaxies. Merger fractions in AGN hosts are mostly found to be consistent with those of matched control galaxies \cite[e.g.][]{kocevski_candels:_2012,villforth_morphologies_2014,villforth_host_2017,mechtley_most_2016, marian_major_2019}, suggesting that mergers are not closely linked to AGN activity. Other studies have found extremely high detected merger fractions in AGN host galaxies, unlikely to be consistent with merger fractions in the general galaxy population \citep{urrutia_evidence_2008,glikman_major_2015,chiaberge_radio_2015}. These seemingly contradictory results have raised the question of how major mergers are linked to AGN triggering and if differences in triggering mechanisms exist between AGN of different luminosities and physical properties.

Some theoretical models have suggested that major mergers become prevalent only at the highest AGN luminosities \citep{hopkins_characteristic_2009,hopkins_we_2013} since the fuel mass for low luminosity AGN can easily be supplied by secular processes, whereas fuel masses for high luminosity AGN exceed the rates of inflow possible in dynamically stable galaxies. Some observational work has found an increase of merger signatures with AGN luminosity in compilations of literature data \citep{treister_heavily_2009, fan_most_2016, glikman_major_2015}. However, several studies of high luminosity AGN found their merger fractions to be relatively low and consistent with those of control galaxies, suggesting that mergers are not strongly connected to even the highest luminosity AGN \citep{mechtley_most_2016,villforth_host_2017,marian_major_2019}.

Theoretical models also suggest a correlation between obscuration and merger incidence \citep[see e.g.][]{sanders_ultraluminous_1988, di_matteo_energy_2005, hopkins_cosmological_2008, alexander_what_2012}:  "young" obscured AGN are predicted to have higher detected merger fractions since they appear closer to the merger, while "old" unobscured AGN appear late and show weaker merger features \citep[e.g.][]{kocevski_are_2015, fan_most_2016, glikman_major_2015}.

A significant amount of work has been done to date analyzing the incidence of merger features in AGN host galaxies across a wide range of AGN luminosities, redshift, and for a wide range of AGN types. However, no complete and systematic analysis of detected merger fractions exists to date. A joint analysis of literature data is made difficult by several factors. The "detected merger fractions" reported in the literature are not a clearly defined quantity. Merger fractions can be measured either quantitatively \citep[e.g.][]{conselice_asymmetry_2000,pawlik_shape_2016} or qualitatively through visual inspection \citep{kartaltepe_candels_2014}. These different methods can yield results that show large discrepancies based on the stage of the merger and merger mass ratio \citep{lotz_effect_2010,lotz_effect_2010-1,pawlik_shape_2016}. Additionally, the data used can cover a wide range of depth, resolution, and rest wavelength, meaning studies do not have the same sensitivity to merger features. Detected merger fractions therefore carry significant systematic uncertainties and potential systematic biases between studies need to be addressed. Despite these difficulties, a systematic analysis is needed first of all to assess if detected merger fractions as reported in the literature are consistent across similar samples and secondly to determine if the wealth of data collected so far shows any evidence of trends with luminosity, redshift or AGN selection methods, as suggested in both theory and previous collection of observational data.

In this paper, I will collect all available literature data to create a complete catalogue of merger fractions in AGN host galaxies. I will analyze if differences reported in the literature are likely due to differences in methodology or physical differences in samples. I will study the incidence of merger features in AGN as a function of bolometric luminosity, redshift, as well as selection method. The collection of data from the literature, as well as the derivation of merger fractions and bolometric luminosities, is explained in Section \ref{S:data}. The results are presented in Section \ref{S:results}. I discuss the limitations of this approach and discrepancies across studies in Section \ref{S:discussion}, followed by conclusions in Section \ref{S:conclusion}. A detailed summary of how the data were extracted for individual studies is given in Appendix \ref{A:studies}.

\begin{figure*}
\begin{center}
\includegraphics[width=18cm]{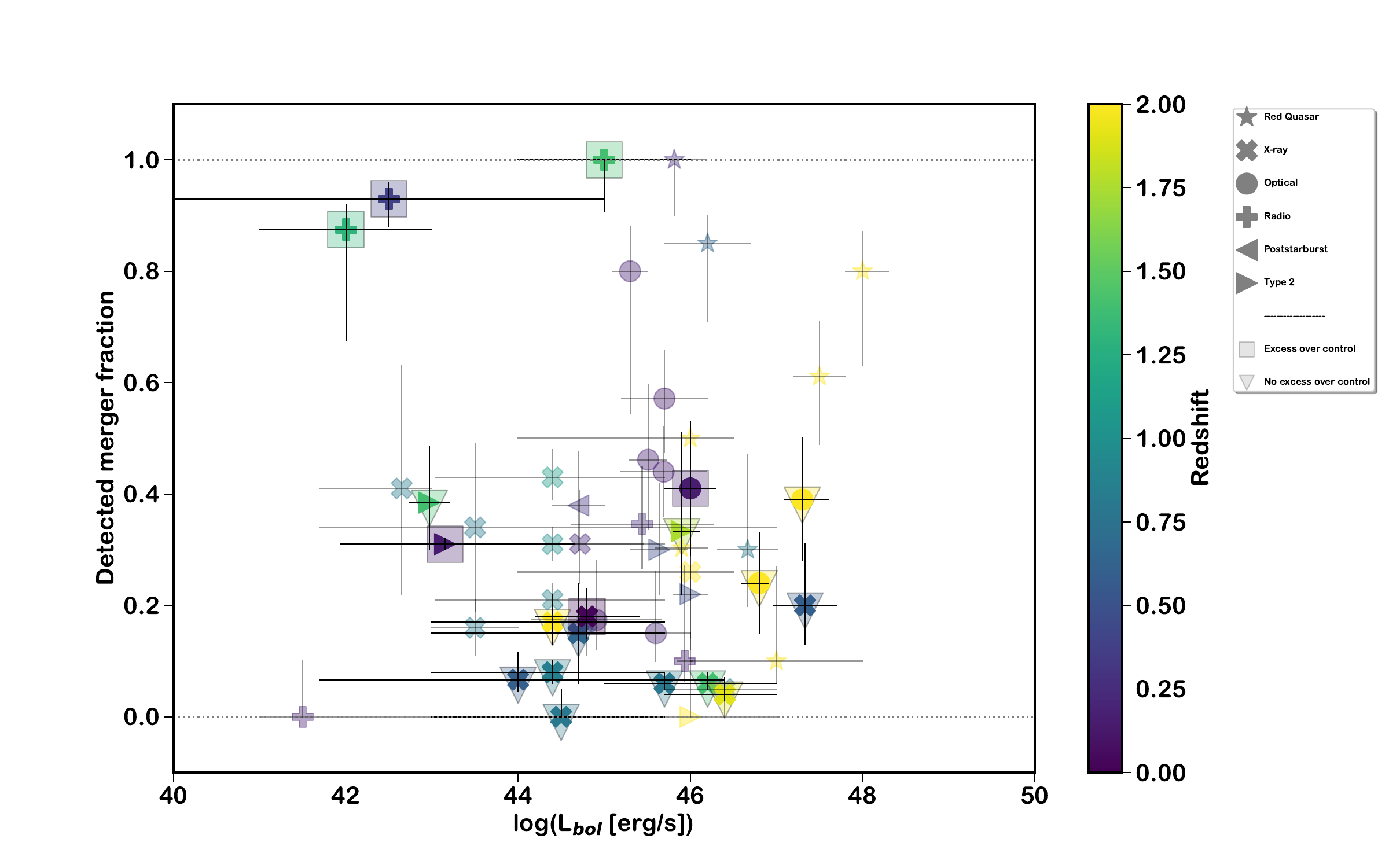}
\caption{The detected merger fraction of AGN as a function of bolometric AGN luminosity. Colour scale indicates the average redshift of the study. Solid coloured symbols have control sample, semi-transparent symbols do not have a control sample. Large squares surrounding a symbol indicate that an excess over control was detected while downward arrows around the symbol indicate that no excess over control at 3$\sigma$ was detected. Symbols indicate the selection method: star: red quasar; filled cross: X-ray; filled circle: optical; filled plus: radio; arrow left: post-starburst; arrow right: Type 2. }
\label{F:mainplot}
\end{center}
\end{figure*}

\section{Data}
\label{S:data}

In this paper, I will collect a complete catalogue of detected merger fractions in the AGN population from the literature and analyze the difference between samples as well as trends with redshift, luminosity, and selection methods. The aim is to collect the data as consistently as possible, calculate bolometric luminosities in a consistent fashion, and create a complete catalogue of merger fractions in AGN host galaxies.

I collect from the literature all studies that report merger fractions in AGN host galaxies. I do not include studies that analyze the incidence of AGN in mergers \citep[e.g.][]{ellison_galaxy_2011,ellison_galaxy_2013,ellison_galaxy_2015,satyapal_galaxy_2014,sabater_triggering_2015} since the enhancement of AGN activity triggered during mergers cannot be translated into merger fractions in AGN samples, but will discuss those results in Section \ref{S:discussion}. I consider all papers included in a previous study with the same aim \citep{treister_major_2012}, and check if the required data is available in the paper. The following papers are considered, listed in alphabetical order:\citet{bahcall_hubble_1997,bennert_evidence_2008,boehm_agn_2012,cales_hubble_2011,canalizo_quasi-stellar_2001, chiaberge_radio_2015, cisternas_bulk_2011,del_moro_mir_2016, donley_evidence_2018, dunlop_quasars_2003, ellison_definitive_2019, fan_most_2016, georgakakis_host_2009, glikman_major_2015, goulding_galaxy_2018, grogin_agn_2005, hewlett_redshift_2017, hong_correlation_2015,hutchings_optical_1984, kocevski_candels:_2012, kocevski_are_2015, koss_merging_2010, kartaltepe_multiwavelength_2010, lanzuisi_compton_2015,liu_host_2009, marian_major_2019, marian_significant_2020, mechtley_most_2016, ramos_almeida_are_2011, schawinski_hst_2011, schawinski_heavily_2012, urrutia_evidence_2008, veilleux_deep_2009, villforth_morphologies_2014, villforth_host_2017, villforth_host_2019, wylezalek_towards_2016,zakamska_host_2019}. To my knowledge, this includes all studies of mergers in AGN host galaxies.

A small number of studies were not included due to unavailability of data \citep[these are:][]{lanzuisi_compton_2015,kartaltepe_multiwavelength_2010, schawinski_hst_2011, schawinski_heavily_2012, ellison_definitive_2019, koss_merging_2010, goulding_galaxy_2018}\footnote{\citet{kartaltepe_multiwavelength_2010},  \citet{schawinski_heavily_2012} and \citet{goulding_galaxy_2018} do not report AGN luminosities for the full samples; \citet{koss_merging_2010} and \citet{ellison_definitive_2019} do not report luminosities for sub-samples; \citet{schawinski_hst_2011} do not report detected merger fractions}, see Appendix \ref{A:studies} for details.

For all suitable studies, I report the detected merger fraction, bolometric luminosity, redshift, and selection method. If control samples are available, I also give the merger fraction in the control sample and calculate if there is a statistically significant excess of mergers in the AGN sample. Details on the data collection are outlined below.

\begin{figure}
\begin{center}
\includegraphics[width=10cm]{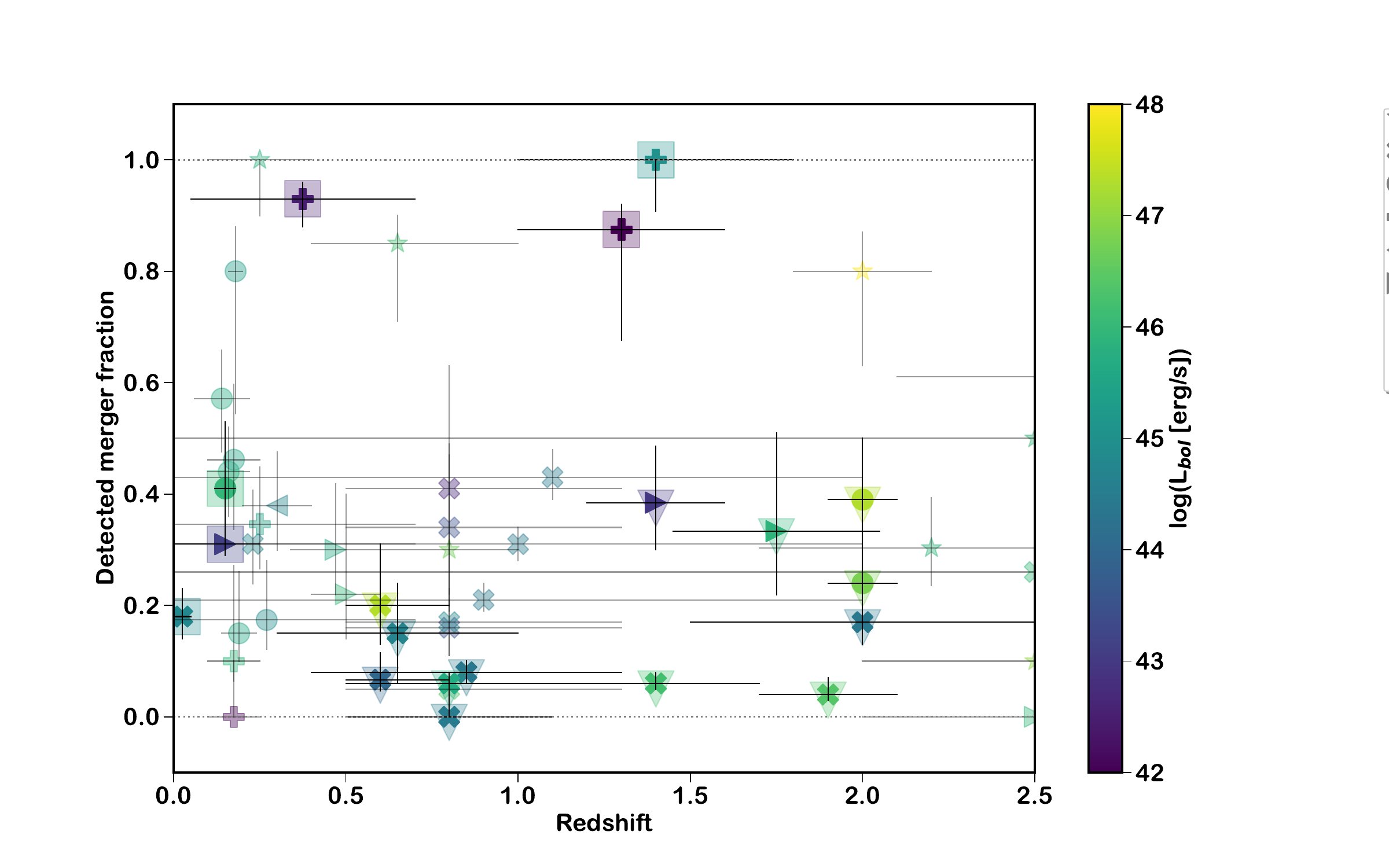}
\caption{The detected merger fraction of AGN as a function of redshift.  Colour scale indicates the average bolometric luminosity of the study. Symbols as in Figure \ref{F:mainplot}.}
\label{F:mainplot_z}
\end{center}
\end{figure}

\subsection{Caveats of a literature review of mergers in AGN hosts}
\label{S:caveats}

Before presenting and analyzing these results, I will discuss the caveats of this approach:

\begin{itemize}
\item Meaning of "merger fraction": the term merger fraction refers to the fraction of sources with detected merger features. This should not be confused with the fraction of objects undergoing a merger. I therefore refer to detected merger fractions, rather than merger fractions throughout the paper. All merger detection methods are subject to both false positives and false negatives. The detection probability for merger depends on the merger mass ratios, viewing angle, time since merger, the wavelength used for imaging, resolution as well as the depth of data \citep{lotz_galaxy_2008,lotz_effect_2010,lotz_effect_2010-1, barnes_transformations_1992, hernquist_structure_1992}. For example, deeper imaging studies will be able to detect fainter merger features, which allows the detection of both lower mass ratio mergers and longer delay after the mergers. Different merger detection methods also have different sensitivities and specificities \citep{lotz_major_2011,villforth_morphologies_2014,pawlik_shape_2016}. The reported merger fractions, therefore, do not directly translate to actual merger fractions. Additionally, due to the rarity of mergers, the contamination in merger samples can be high. I will address this issue when comparing results across different studies in Sections \ref{S:results} and \ref{S:discussion}.
\item Control samples: Mergers occur in the general galaxy population and the merger fractions depend on the galaxy mass, mass ratio, and redshift \citep[see e.g.][]{lotz_major_2011,lacey_merger_1993,hopkins_we_2013}. If (actual or detected) merger rates in AGN samples match those of control galaxies, this implies that AGN are not in fact triggered by mergers. An excess in detected merger fractions needs to be present to indicate a connection between mergers and AGN. When available, detected merger fractions in control samples are compared to the detected merger fractions in AGN host galaxies. Due to potentially high contamination in merger samples, a comparison of merger rates between AGN and controls can be biased \citep{lambrides_merger_2021}.
\item Redshift evolution: the samples included span a wide range of redshifts. The properties of galaxies change considerably through cosmic time. Merger rates evolve significantly, decreasing at lower redshift \citep{lotz_major_2011,hopkins_we_2013}. Additionally, high redshift galaxies are more gas-rich \citep[e.g.][]{genzel_combined_2015}, although the extent of this increase is still under debate \citep{narayanan_galaxy_2012}. Since gas-rich mergers show more pronounced merger signatures \citep{lotz_effect_2010}, this is expected to affect the detected merger fraction. Additionally, gas-rich galaxies can show clumps and other asymmetric features that could be wrongly identified as mergers \citep{bournaud_observations_2008}.
Therefore, samples across different redshifts cannot be easily compared due to changes in the underlying galaxy population. I will analyze the detected merger fraction as a function of redshift in Section \ref{S:results}.
\end{itemize}

These caveats need to be kept in mind when comparing detected merger fractions across samples and will be discussed in detail in Section \ref{S:discussion}.


\begin{figure*}
\begin{center}
\includegraphics[width=8cm]{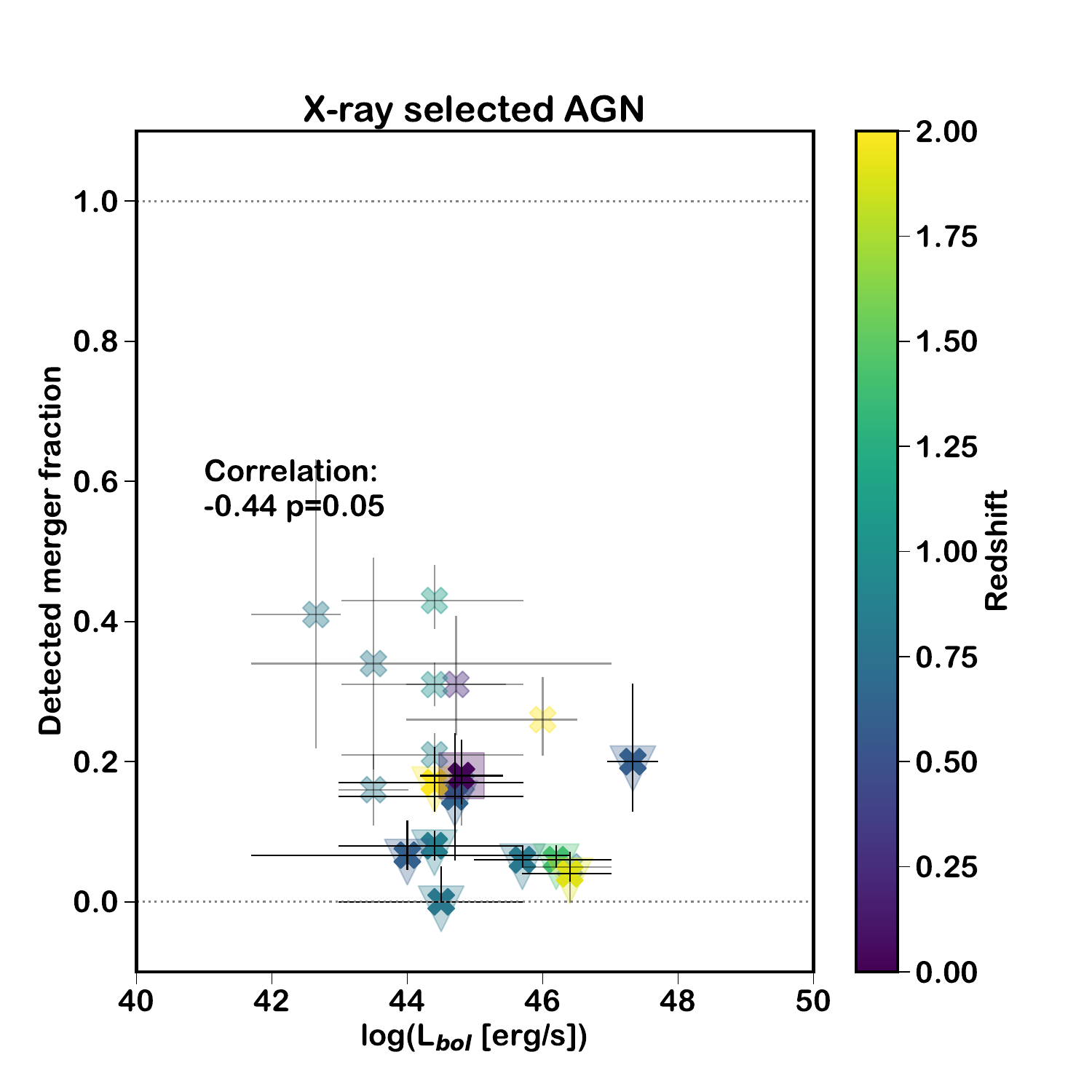}
\includegraphics[width=8cm]{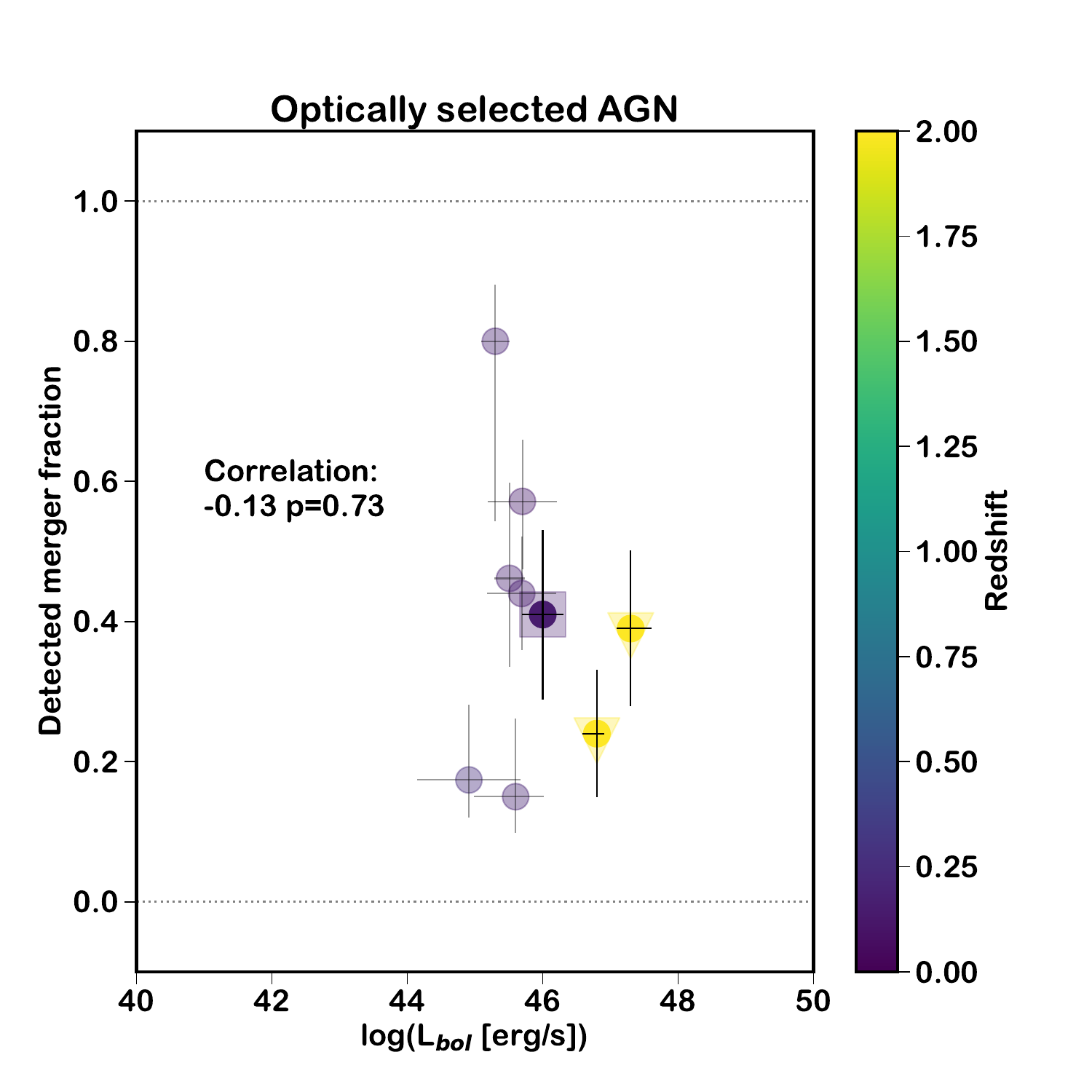}
\includegraphics[width=8cm]{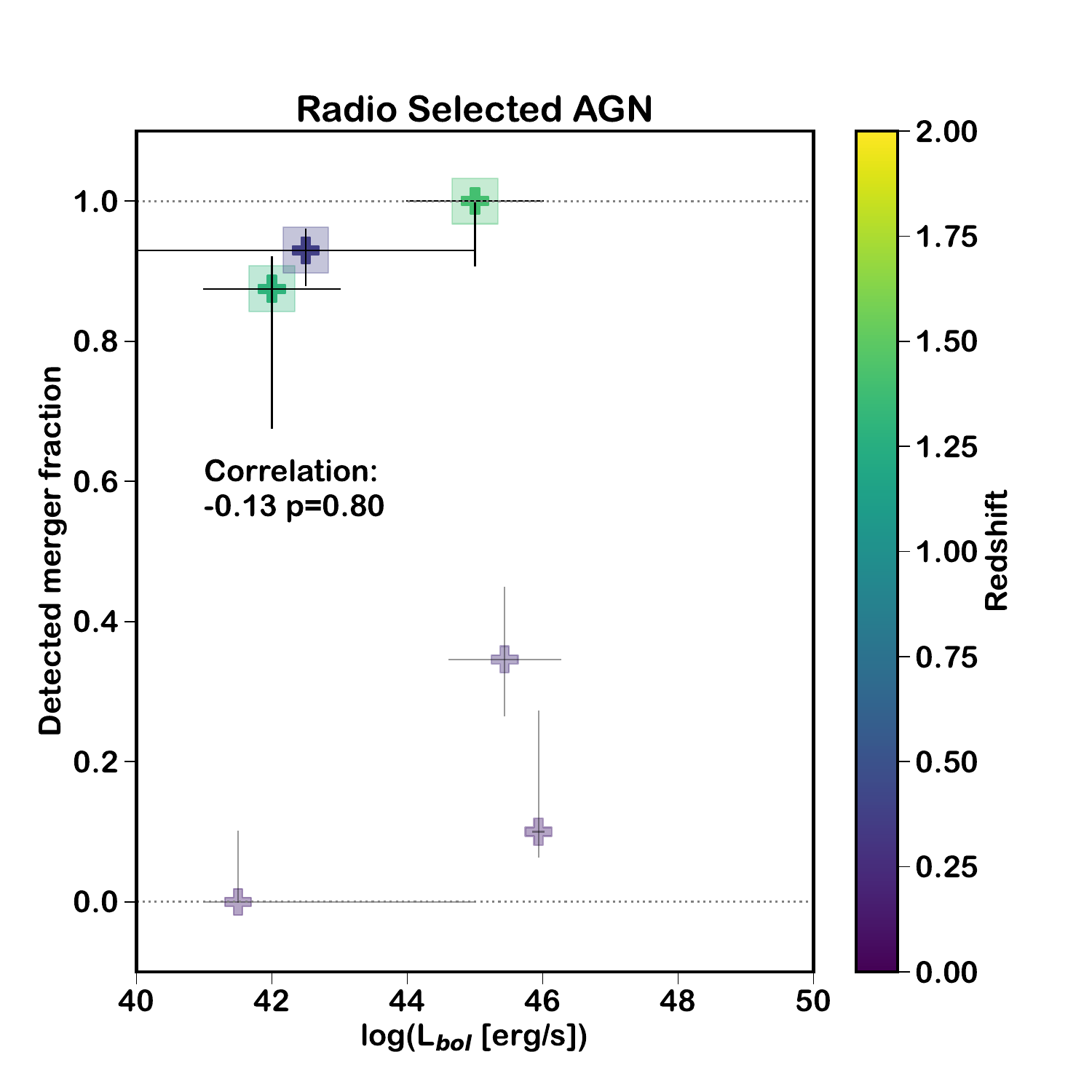}
\includegraphics[width=8cm]{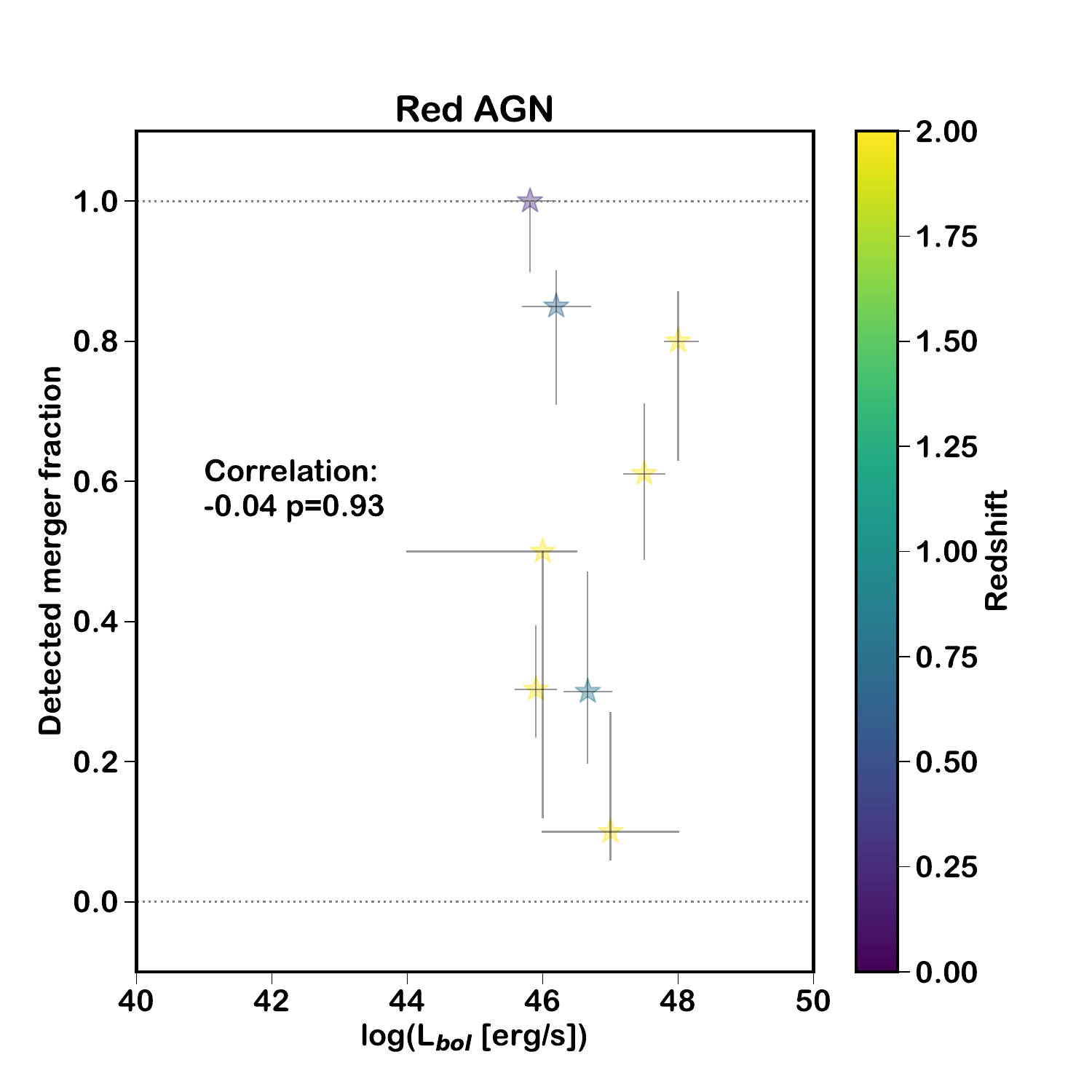}
\includegraphics[width=8cm]{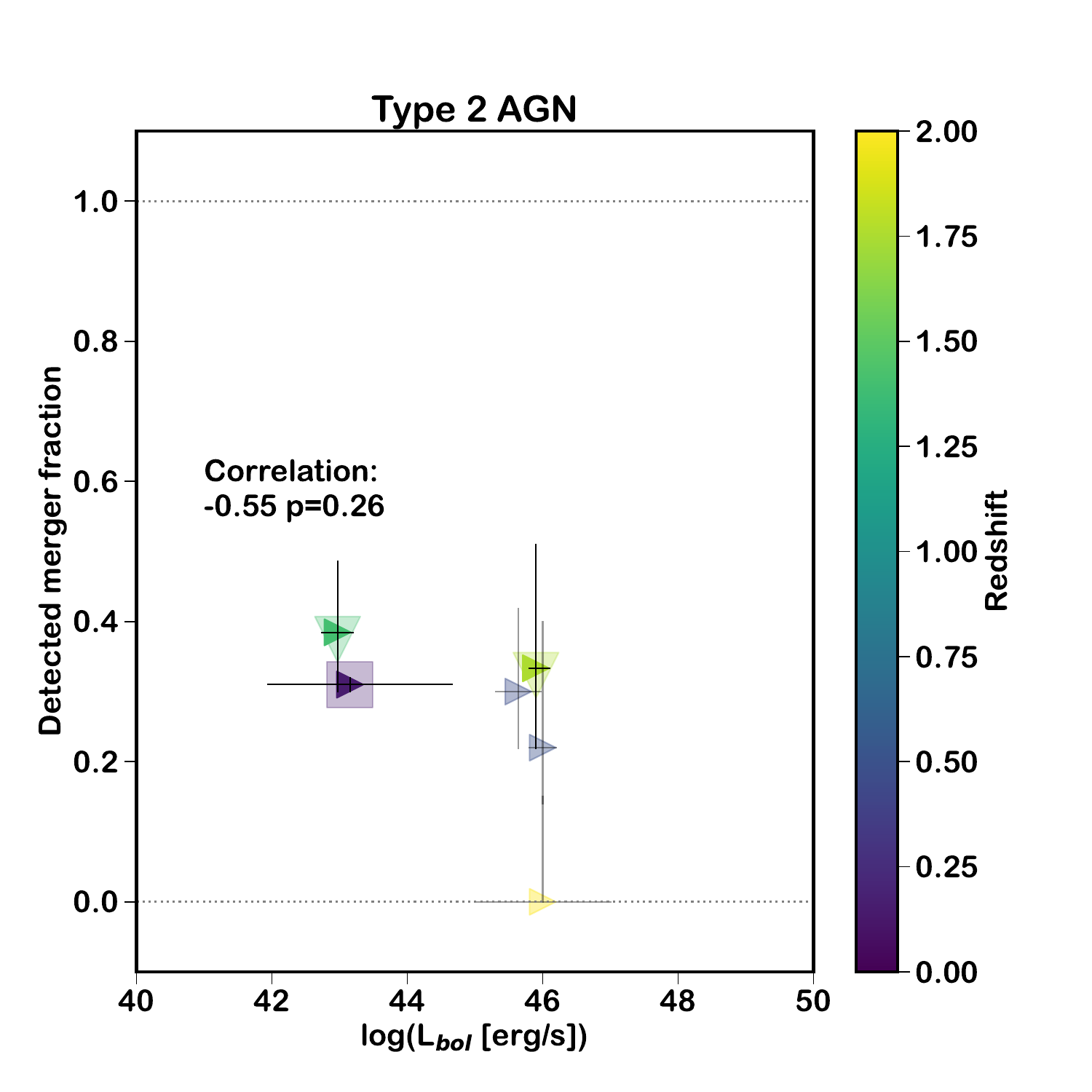}
\includegraphics[width=8cm]{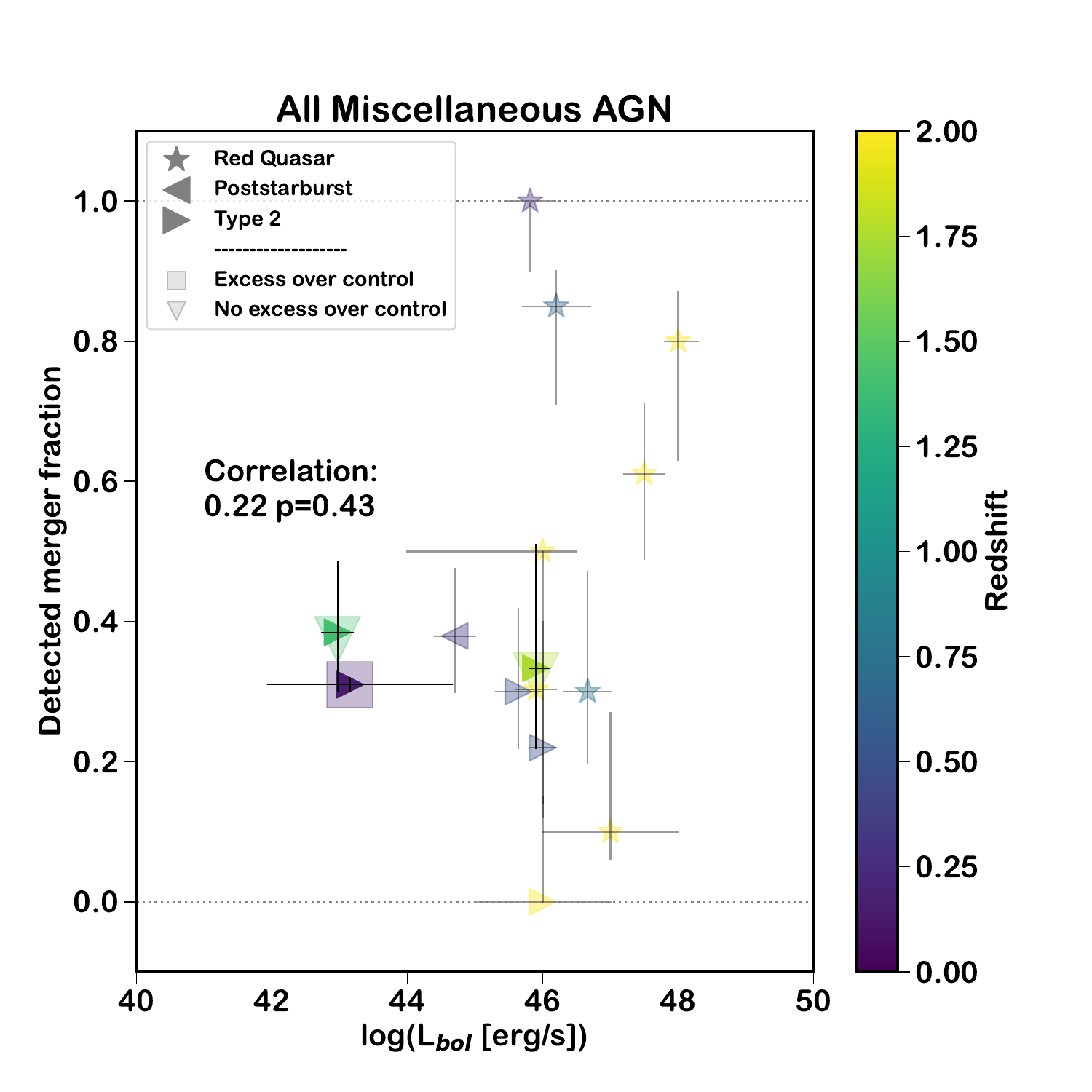}
\caption{Same as Figure \ref{F:mainplot}, but showing results separately for different selection methods, specifically, from top left to bottom right: X-ray selected, optically selected, Radio selected, red quasars, type 2 AGN and miscellaneous methods.}
\label{F:mainplot_method}
\end{center}
\end{figure*}

\subsection{Merger fractions}
\label{S:f_merg}

Detected merger fractions for all studies are extracted following the general approach below to derive the fraction of AGN currently undergoing major mergers. Specifically:

\begin{itemize}
\item The aim is to select gravitationally strongly interacting objects that are near or post-coalescence. Objects with nearby or non-interacting neighbours are therefore not included to avoid contamination by galaxies in dense environments. 
\item If different merger classifications are available and example images are included, I use those to identify the closest matching category.
\item If different merger classifications are available, but no example images are available, I choose the closest matching category from the description in the text.
\item If only a single merger fraction is reported, I use this value. I do not re-analyze datasets or perform visual inspections.
\item Any cases that do not match these rules are decided on an individual basis, see Appendix \ref{A:studies}.
\end{itemize}

The aim of this approach is to select ongoing or completed, major mergers (mass ratios $\gtrapprox 1/4$). While the above approach does not correct for differences in merger detection, it minimizes bias due to differences in visual classifications. In the remainder of the paper, I will refer to detected major merger fractions, although no clear mass ratio cut-off is made.  The caveats when interpreting detected merger fractions are discussed in more detail in Section \ref{S:caveats}). For details on individual studies, see Appendix \ref{A:studies}.

\subsection{Bolometric corrections}
\label{S:bolcor}

Bolometric luminosities require correcting the luminosities in a given waveband with a bolometric correction. This, therefore, assumes a specific spectral energy distribution. To limit biases, I, therefore, aim to apply bolometric corrections based on the same spectral energy distribution (SED), minimizing biases.  For all studies in the sample, bolometric luminosities are calculated as follows:

\begin{itemize}
\item Optical: For optical luminosities or magnitudes, I use the following bolometric correction from \citet[][p.157, eq. 7.3]{netzer_physics_2013} 
\begin{equation}
\textrm{BC}_{5100\AA} = 53 - \textrm{log}(L_{5100\AA})
\end{equation}
\label{EQ:lbol_5100}
where $BC_{5100\AA}$ is the bolometric correction and $L_{5100\AA}$ is the luminosity ($\lambda L_{\lambda}$ or $\nu L_{\nu}$). Any corrections to the luminosity due to different wavelength are discussed for the individual cases, but in most cases, the fluxes are close enough in wavelength that the above relation is valid. Whenever possible, I use AGN rather than overall luminosities (i.e. from image decomposition).
\item X-ray: For X-ray, I use  he following correction to convert the X-ray to optical luminosities from \citet[][p.157, eq. 7.5]{netzer_physics_2013} 
\begin{equation}
\textrm{log}(L_{5100\AA}) = 1.4 \times \textrm{log}(L_X) - 16.8
\end{equation}
\label{EQ:lbol_x}
where $L_X$ i the luminosity at 2-10kev. Due to the fact that generally, no spectral information is available, no k-correction is applied.
\item \textrm{[OIII]}: for $L_{\textrm{[OIII]}}$, I use the luminosity dependent bolometric corrections from \citep{lamastra_bolometric_2009}.
\item Radio: Radio luminosities have the largest uncertainties in the bolometric corrections due to the large difference in AGN SEDs in this wave-length regime. Therefore, I take the following conservative approach to take this uncertainty into account: first, I use the relation between Radio and [OIII] luminosity from \citep{best_fundamental_2012} to estimate the expected range in [OIII] luminosities. I then follow the approach for the [OIII] luminosities (see above). The uncertainties in bolometric luminosities from radio luminosities are high, this reflects the uncertainty in the bolometric corrections applied.
\item Other: for studies that either give their own bolometric luminosities or do not fall into the above categories, a decision is made on an individual basis. See Appendix \ref{A:studies} for details.
\end{itemize}

Therefore, consistent bolometric corrections are applied across the sample. Samples with similar selection methods will have the same bolometric correction applied, further minimizing any potential bias. For details on how bolometric correction are applied in individual studies, see Appendix \ref{A:studies}.

\subsection{AGN selection method}
\label{S:data_selection}

The selection method for each AGN sample is reported. When the selection method used is unclear, I discuss the adopted selection method in Section \ref{A:studies}. The different selection method categories used here are:
\begin{itemize}
\item X-ray: selected in the hard or soft X-rays, bolometric corrections are applied following Section \ref{S:bolcor}.
\item Optical: selected in the optical to have a dominant blue continuum (i.e. Type 1 AGN), bolometric corrections are applied following Section \ref{S:bolcor}.
\item Radio: selected in radio bands, identification of the AGN nature in some cases confirmed in the optical, bolometric correction applied following the procedure outlined in Section \ref{S:bolcor}
\item Type 2: identified as an AGN using narrow emission line diagnostics, bolometric corrections are applied following procedure in Section \ref{S:bolcor}
\item red AGN: contains samples selected in the IR or showing strong reddening. This category contains AGN selected using IR power-law techniques \citep[e.g.][]{donley_evidence_2018}, radio pre-selection \citep[e.g.][]{urrutia_evidence_2008} as well as AGN with SEDs showing signs of strong obscuration \citep[e.g.][]{villforth_host_2019,zakamska_host_2019}. See the Appendix for details and how bolometric corrections are applied.
\item Miscellaneous: Additionally, one sample \citep{cales_hubble_2011} does not fit any of the above categories and is listed as post-starburst. 
\end{itemize}

\subsection{Other general rules for data collection}
\label{S:data_misc}

The following rules highlight other general approaches taken for the collection of data. See Appendix \ref{A:studies} for details on individual studies.

\begin{itemize}
\item Detected merger fractions and uncertainties are calculated following \citet{cameron_estimation_2011}. Errors reported are 1$\sigma$ uncertainties. For studies including control samples, merger fractions in the control sample are calculated using the same method. I determine if any sample shows an excess above 1-tailed 3$\sigma$ (p=0.003) significance. The probability of detected merger fractions being consistent is calculated by multiplying the respective probability distributions.
\item Sub-samples for each study are included separately, when available.
\item For both redshift and luminosity, whenever possible, I extract the data from the paper to calculate the mean and standard deviation. If this is not possible and only general sample properties are given, I give the ranges for those values.
\item  The telescope and filter used for observations as well as the rest-frame wavelength of observations are reported.
\item Rejected studies: for all studies, I attempt collect the relevant information, also taking into account papers referenced  for sample selection. If it is not possible to extract all relevant data, I do not include the sample. This is the case for either entire or sub-samples of \citet{kartaltepe_multiwavelength_2010, schawinski_hst_2011, schawinski_heavily_2012, ellison_definitive_2019, koss_merging_2010}, mostly due to AGN luminosities not being available. Detailed descriptions for those cases are given in Appendix \ref{A:studies}.
\end{itemize}

\section{Results}
\label{S:results}


I present a complete catalogue of detected merger fractions in AGN host galaxies, compiled as described in Section \ref{S:data}. All results can be found in Table \ref{maintable}, detailed comments on how data are extracted from each paper are given in Appendix \ref{A:studies}.

This complete catalogue includes data from 33 papers, with 50 separate samples, covering all reported detected merger fractions in AGN host galaxies since 1984. The redshifts of the samples studied have a mean and standard deviation of 0.98$\pm$ 0.79  and a range of $0.025 \leq z \leq 2.0$. The bolometric luminosities of all samples have a mean of $\textrm{log}(L_{\textrm{log}} [erg/s]) = 45.2 \pm 1.4$ and a range of $41.5 \leq \textrm{log}(L_{\textrm{log}} [erg/s]) \leq 48.0$. The range of expected merger fraction spans all the way from 0 to 1. The detected merger fractions have a mean and standard deviation of 0.34 $\pm$ 0.27. Note that these numbers give the average over samples of AGN, rather than objects. 19 of our 50 samples have a control sample, of those 6 (32\%) have detected an excess of detected merger fraction of at least 3$\sigma$ significance. The number of studies showing an excess above control merger fractions is well in excess of expected false positive rate.

Figure \ref{F:mainplot} shows the detected merger fraction for all collected studies as a function of bolometric luminosity. The majority of AGN samples have low detected merger fraction ($<$20\%). Once all literature data is taken into account, the detected merger fraction in AGN host galaxies shows no correlation with luminosity, in contradiction to previous work comparing smaller datasets \citep[][see also Table \ref{T:correlation} for correlation coefficients]{treister_major_2012,glikman_major_2015, fan_most_2016}. Similarly, there is no clear trend with redshift (Fig. \ref{F:mainplot_z}). Additionally, there is also no correlation between the merger rates and bolometric luminosity when subdividing the sample in redshift. The lack of correlation with redshift suggest that any redshift evolution in the intrinsic merger fraction \citep[e.g.][]{hopkins_we_2013} is compensated either by surface brightness dimming or changes in the host galaxy population properties. Detected merger fractions span a wide range, indicating either large discrepancies in merger incidence between samples or significant systematic uncertainties in the detected merger fractions.

\begin{table}
\centering
\caption{Pearson correlation coefficient for the detected merger rates presented in this sample as a function of both bolometric luminosity and redshift.}
\label{T:correlation}
\begin{tabular}{llcc} 
\hline
Sample & Data &  Correlation Coefficient $\rho$ & p-value\\
\hline
All & $f_\textrm{m}$ vs L$_{\textrm{bol}}$ & -0.02 & 0.91\\ 
All & $f_\textrm{m}$ vs z & -0.03 & 0.83 \\ 
\hline
Radio & $f_\textrm{m}$ vs L$_{\textrm{bol}}$ & -0.13 & 0.80 \\
X-ray & $f_\textrm{m}$ vs L$_{\textrm{bol}}$ & -0.44 & 0.05 \\
Optical & $f_\textrm{m}$ vs L$_{\textrm{bol}}$ & -0.13 & 0.76 \\
Type 2 & $f_\textrm{m}$ vs L$_{\textrm{bol}}$ & -0.61 & 0.27 \\
Red and IR AGN & $f_\textrm{m}$ vs L$_{\textrm{bol}}$ & -0.18 & 0.70 \\
All Misc & $f_\textrm{m}$ vs L$_{\textrm{bol}}$ & 0.22 & 0.47 \\
\hline
\end{tabular}
\end{table}

I will now compare results by AGN selection method. AGN samples selected using the same method have similar selection effects. Since the selection wavelength is used to calculate bolometric luminosities (see Section \ref{S:bolcor}), systematic uncertainties in bolometric corrections are minimized. Due to the similarities in the spectral energy distribution, AGN selected using the same method will also show similar contrast between AGN and host galaxy. This will allow  to more clearly assess if differences between samples suggest large differences in the methodology and data or intrinsic differences between samples.

X-ray selection is the most common selection method across samples, with 20 samples in this category. X-ray selected samples are often conducted in deep fields with deep high resolution imaging  \citep[e.g.][using CANDELS, GOODS and COSMOS data]{boehm_agn_2012,georgakakis_host_2009,kocevski_candels:_2012,villforth_morphologies_2014}. Despite often having similar luminosity ranges and data sets, these samples show a wide range in detected merger fractions, from $\sim$ 0-40\% with a mean of 18\%. The scatter in the results is however consistent with the uncertainties in individual samples (mean of the standard deviations $<\sigma> = 0.11$, standard deviation of mean $\sigma_{\mu} = 0.12$). Due to the wide availability of control samples in deep fields, many X-ray studies have control samples. In all deep field samples, these studies find no excess over control \citep{grogin_agn_2005, cisternas_bulk_2011,boehm_agn_2012,kocevski_candels:_2012,hewlett_redshift_2017}. Similarly, \citet{villforth_host_2017} find no excess over control for an X-ray selected sample aimed to extend such studies to higher luminosities. The only X-ray selected sample with an excess of detected merger fractions over control is that by \citet{koss_merging_2010}. This is a low redshift (z$<$0.05) AGN sample using Sloan Digital Sky Survey (SDSS) imaging. The sample size in \citet{koss_merging_2010} is smaller than that in other X-ray selected samples with a matched control, indicating that the detection of an excess is not due to better statistics. The detected merger fraction in this sample is 18 times higher than that in the control sample, well in excess of enhancements of ~2-5 found in other work \citep[e.g.][]{ellison_definitive_2019}. The \citep{koss_merging_2010} sample therefore does not match other X-ray detected samples. This could either be due to the lower redshift, differences in image quality or physical differences between the sample. Due to the fact that \citep{koss_merging_2010} use colour SDSS images, the surface brightness limit cannot be easily compared to the Hubble Space Telescope (HST) imaging data used for other X-ray samples. The \citet{koss_merging_2010} sample was selected in the hard X-rays, thereby favouring potentially more heavily obscured AGN.

A negative correlation between luminosity and detected merger fraction is detected in the full X-ray selected sample, although it is only marginally significant (p=0.05). This could be due to less favourable contrast at higher luminosities. The detected merger fractions in X-ray selected AGN are therefore broadly consistent across studies \citep[besides the low redshift sample by][]{koss_merging_2010}. The data show that X-ray selected AGN on a whole are not strongly associated with recent major mergers.

Nine AGN samples are classed as optically selected, they show a wide range of detected merger fractions from 15-80\% with a mean of 40\%, almost double that in the X-ray selected sample. The scatter between studies is consistent with the uncertainties (mean of the standard deviations $<\sigma> = 0.21$, standard deviation of mean $\sigma_{\mu} = 0.19$). At high redshift ($z>2$), \citet{mechtley_most_2016} and \citet{marian_major_2019} both show detected merger fractions $\sim$20-40\%, although neither find an excess over control samples. Both the detected merger fraction and lack of excess over control match X-ray selected AGN at similar luminosities \citep{villforth_host_2017}. The seven low redshift ($z \leqslant 0.7$) optical AGN samples show a range of detected merger fractions $\sim$15-80\% \citep{bahcall_hubble_1997,hong_correlation_2015, hutchings_optical_1984,dunlop_quasars_2003,
veilleux_deep_2009,bennert_evidence_2008, marian_significant_2020}. These detected merger fractions are in excess of the X-ray detected fractions at similar luminosity and redshift. The only low redshift optically selected samples with a control sample shows an excess in the detected merger fraction \citep{marian_significant_2020}.  \citet{bennert_evidence_2008} found the highest detected merger fraction of $\sim$80\% in optically selected AGN, likely due to significantly deeper imaging. \citet{veilleux_deep_2009} study optically selected AGN with a range of FIR excesses and find a detected merger fraction of 57\%. The FIR excess implies some of their sources are likely to be associated with stronger starbursts, they find higher detected merger fractions in FIR strong sources. It is therefore unclear if the differences seen reflect high detected merger fractions in some samples or differences in data quality and selection method.

Radio detected and red AGN both show a much more significant spread in detected merger fractions. In both categories, the scatter in detected merger fractions seen between studies is well in excess of what is expected from the uncertainties in the detected merger fractions (mean of the standard deviations $<\sigma> = 0.15/0.15$, standard deviation of mean $\sigma_{\mu} = 0.41/0.29$ for radio and red samples respectively). Both samples also show a a large proportion of detected merger fractions well in excess of 50\%.
  
Starting with the radio-selected AGN, high detected merger fractions are reported in samples by \citet{ramos_almeida_are_2011} and \citet{chiaberge_radio_2015}, whereas very low detected merger fractions are observed by \citet{dunlop_quasars_2003} and \citet{hutchings_optical_1984}, despite the samples being of similar redshift. While the samples from \citet{chiaberge_radio_2015} are at moderate to high redshift ($\sim$1.5), the other merger excess sample \citep{ramos_almeida_are_2011} is at lower redshift ($0.05<z<0.7$). The radio samples with low detected merger  fractions \citep{dunlop_quasars_2003,hutchings_optical_1984} are at similarly low redshift ($0.05<z<0.7$). All samples span wide luminosity ranges ($41 < \textrm{log}(L_{\textrm{bol}} [erg/s]< 45)$), this wide range in luminosity is at least partially due to the fact that bolometric corrections for radio samples carry the largest uncertainties (see Section \ref{S:bolcor}). Another difference between radio selected samples is in the image quality, there are mixture of space based \citep{chiaberge_radio_2015,dunlop_quasars_2003} and ground-based data \citep{hutchings_optical_1984,ramos_almeida_are_2011}, meaning differences in spatial resolution and likely surface brightness limits. Additionally, earlier studies \citet{hutchings_optical_1984,dunlop_quasars_2003}, might have required confirmation in the optical (this is not entirely clear from the sample selection sections in those papers), while \citet{chiaberge_radio_2015,ramos_almeida_are_2011} do not. This would mean that samples by \citet{hutchings_optical_1984,dunlop_quasars_2003} are in fact more similar to optically selected samples, for which they would fit in the extremely wide range of detected merger fractions observed. The differences seen could therefore reflect significant differences in data quality, as well as potentially differences in the selection method and therefore SED .

Red AGN also show a very wide range in detected merger fractions, from $\sim$10-100\%, with more than half in excess of 50\%. All red AGN samples have comparatively high luminosity ($46 < \textrm{log}(L_{\textrm{log}} [erg/s]< 48)$). The scatter in detected merger fractions is in excess of what is expected from the uncertainties in individual studies (mean of the standard deviations $<\sigma> = 0.14$, standard deviation of mean $\sigma_{\mu} = 0.29$). The sample by \citet{urrutia_evidence_2008} was pre-selected in the radio and shows an extremely high detected merger fraction of 85\%. On the other hand, the Low Ionization Iron Broad Absorption Line (FeLoBAL) sample by \citet{villforth_host_2019} in the same luminosity range has a relatively low detected merger fraction of 30\%.  While the \citet{villforth_host_2019} FeLoBAL sample has no galaxy control sample, its merger rate, luminosity, redshift as well as data quality and methodology matches that of the X-ray selected sample by \citet{villforth_host_2017}, which shows no excess over control. FeLoBALs show connections to red quasar samples, the sample from \citet{urrutia_evidence_2008} also shows very high incidences of FeLoBALs. FeLoBALs are also known to show high levels of reddening and obscuration \citep{dai_intrinsic_2012,dunn_determining_2015}. Similar to the \citet{villforth_host_2019} FeLoBAL sample, the extremely red quasar sample from \citet{zakamska_host_2019} shows low detected merger fractions of 10\%. \citet{del_moro_mir_2016} studied a sample of mid-IR luminous AGN, of which 24-48 \% were found to be compton thick. The detected merger fraction across the sample is 30\%, with a weak trend for a higher disturbed fraction in the more X-ray obscured sources. In summary, while some red and IR selected quasars show extremely high detected merger fractions \citep{canalizo_quasi-stellar_2001,urrutia_evidence_2008,glikman_major_2015, fan_most_2016}, this is not universally true for red and IR selected AGN \citep{villforth_host_2019, zakamska_host_2019}. The interesting differences between high and low detected merger rate samples here is that the samples from \citet{urrutia_evidence_2008} and \citet{glikman_major_2015} with high detected merger fractions were initially radio-selected, whereas those from \citet{villforth_host_2019} and \citet{zakamska_host_2019} are selected from the SDSS Quasar sample, meaning, they generally require pre-selection in the optical or X-ray. This pattern of optically selected sources having lower detected merger fractions likely matches that seen in the radio selected samples. The 'red quasar' category shows maybe the most wide-ranging set of selection methods, from IR selection \citep{canalizo_quasi-stellar_2001,del_moro_mir_2016,fan_most_2016,donley_evidence_2018} to radio selection combined with extreme levels of obscuration \citep{urrutia_evidence_2008,glikman_major_2015} and SDSS samples with high levels of reddening \citep{villforth_host_2019,zakamska_host_2019}. This could explain the wide discrepancy between different red AGN samples.

There are six samples of Type 2 AGN. The standard deviation of the means of the detected merger fractions (0.13) are comparable to the the standard deviations (0.18).  The Type 2 samples span a very wide range of luminosities ($42 < \textrm{log}(L_{\textrm{log}} \textrm{[erg/s]}< 46)$ and high redshifts $0 < z < 2$). Type 2 AGN with a wide range of luminosities at z$\sim$1.5 from \citep{chiaberge_radio_2015} show no excess over control.  \citet{ellison_definitive_2019} find an excess over control at a comparable luminosity ($\textrm{log}(L_{\textrm{log}} \textrm{[erg/s]} \sim 43)$), but lower redshift (z=0.1). In this case the difference can be easily explained, the samples from \citep{chiaberge_radio_2015} are small ($\sim$ 10), excesses are found, but are not statistically significant (at $\sim$ 1 -2 $\sigma$), while the sample size of \citep{ellison_definitive_2019} is the largest of all studies collected ($\sim$ 1300).  The low redshift Type 2 samples of both \citet{liu_host_2009} and \citet{wylezalek_towards_2016} without a comparison sample also show comparable detected merger fractions to \citet{ellison_definitive_2019}.  The results from these studies are broadly consistent with an enhancement of merger fractions in Type 2 AGN in large samples \citep{ellison_definitive_2019}, but comparably low fractions of these AGN directly associated with mergers. 

Finally, there is one post-starburst \citep{cales_hubble_2011} sample that does not match any category clearly, its detected merger fraction is comparable to AGN of comparable luminosity \citep{liu_host_2009,chiaberge_radio_2015}.

It is therefore clear that while a wide range of detected merger fractions exist in AGN samples, high incidences of merger features are limited to a small set of studies. Looking in more detail at samples with high detected merger fractions ($\geq$ 50\%), three such samples have control samples and all show an excess over control. All these samples are radio selected and have low to moderate bolometric luminosities ($\textrm{log}(L_{\textrm{log}}\textrm{[erg/s]}) < 45$) (\cite{ramos_almeida_are_2011} as well as radio samples from \cite{chiaberge_radio_2015}). Samples with high detected merger fractions but no control sample are red quasar samples at high luminosities ($\textrm{log}(L_{\textrm{bol}}\textrm{[erg/s]}) < 45$), \citep{canalizo_quasi-stellar_2001,urrutia_evidence_2008,glikman_major_2015} as well as some low-redshift optically selected AGN \citep{bennert_evidence_2008,veilleux_deep_2009}. Samples with very high detected merger fractions are therefore predominantly radio detected, red quasars, or optically selected AGN at low redshift. Samples that show an excess of mergers in the AGN sample but moderate detected merger fractions ($\sim$20-50\%) are either moderate-low luminosity Type 2 sample from \citep{ellison_definitive_2019} and the hard X-ray selected BAT sample by \citep{koss_merging_2010}. The majority of samples have detected merger fractions in the 0-30\% range, in many cases, these merger rates are consistent with control sample.

In summary,  in X-ray and optically selected samples, the discrepancies between studies seen in the detected merger fractions are consistent with the uncertainties in measurements. For both selection methods, detected merger fractions are consistent with control samples in all but one sample. Larger discrepancies between studies are seen in radio selected and red quasar samples, where detected merger fractions are generally hogh and in excess of control. For the red quasars, those with radio pre-selection show excess over control, while optically pre-selected ones do not. Type 2 AGN show a mild excess in merger detection over control, although this only becomes apparent in large samples.

Looking at the complete sample of AGN host galaxies, the most notable finding is that there is no correlation in detected merger fraction with luminosity, either in the full sample or when considering only AGN with similar selection methods. As discussed above, there is no indication that systematic biases in the data would mask  a correlation with luminosity. This finding strongly contradicts previous studies analyzing detected merger fractions for smaller sets of studies \citep{treister_heavily_2009,glikman_major_2015,fan_most_2016}. 

\section{Discussion}
\label{S:discussion}

Here, I have analyzed the detected merger fractions as a function of luminosity, redshift, and selection method. No trend is observed between the detected merger fraction and luminosity, in contradiction to previous work analyzing smaller sets of data \citep{treister_major_2012,glikman_major_2015, fan_most_2016}. High detected merger fractions are primarily observed in radio selected and red AGN samples, suggesting potential differences in the merger incidence in those sources. 

In this Section, I will discuss how the detected merger fractions relate to AGN triggering (Section \ref{S:discussion_mergerrates}), if the combined results are consistent with theoretical models (Section \ref{S:discussion_theory}) and how the differences between samples selected in different ways can be explained (Section \ref{S:discussion_singlepop}).

\subsection{Constraints on mergers triggered AGN}
\label{S:discussion_mergerrates}


I will start by recapping the calculation of the contribution of mergers to the AGN population from \citet{villforth_host_2017}. One can calculate the detected merger fraction $r_{\textrm{merger}}$ assuming a duty cycle of AGN $d_{\textrm{merger}}$ during a merger as well as the fraction of galaxies experiencing merger $f_{\textrm{merger}}$. Additionally, we need to consider other potential triggering events such as minor mergers, disk instabilities, or secular processes. The probability of a galaxy currently experiencing such a trigger is given as $f_t$. The duty cycle of AGN in the galaxies experiencing this trigger is given as  $d_t$.  The fraction of mergers in the AGN sample is then given as:

\begin{equation}
\label{eqn:mergAGN}
r_{\textrm{merger, AGN}} = \dfrac{ d_{\textrm{merger}} f_{\textrm{merger}} }{ \sum d_t \times f_t}
\end{equation}

where $r_{\textrm{merger, AGN}}$ is the fraction of (actual, rather than detected) mergers in AGN. Similarly, the rate of mergers in the control sample is:

\begin{equation}
\label{eqn:ctrAGN}
r_{\textrm{merger, Control}} = \dfrac{(1-d_{\textrm{AGN}}) f_{\textrm{merger}}}{f_{\textrm{no\ merger}} + \sum (1-d_t) \times f_t}
\end{equation}

The detected merger fraction further depends on the fraction of mergers and non-merger correctly identified as such, i.e. the true positive rate as well as the false positive rate ($TP/FP$):

\begin{equation}
\label{eqn:corrmerg}
r_{\textrm{merger, detected}} = TP \times r_{\textrm{merger}} \\ + FP \times (1-r_{\textrm{merger}})
\end{equation}

The intrinsic real detected merger fraction therefore depends on the fraction of both mergers and non-mergers identified correctly as well as the intrinsic fraction of mergers. Since mergers are rare, the false positive mergers (second term in equation (5)) can dominate the mergers sample since $r_{merger} << 1$. If the merger rates in the AGN and control sample do not match, the detected excess merger rate depends can be significantly lower than the actual excess \citep{lambrides_merger_2021}.

Most samples with low detected merger fractions are consistent with an AGN duty cycle that is similar in mergers compared to other possible triggers (see Equations \ref{eqn:mergAGN},\ref{eqn:ctrAGN}). Studies that show an excess of AGN in mergers but low merger rates \citep{koss_merging_2010,ellison_definitive_2019} indicate that the duty cycles of AGN are enhanced in mergers, but that the AGN population is not dominated by mergers. Unless the duty cycles of AGN in major mergers are extremely high compared to other triggers, the more common triggers will dominate over rare major mergers, see Eq. \ref{eqn:mergAGN}.

Samples that show high detected merger fractions in excess of control \citep{ramos_almeida_are_2011, chiaberge_radio_2015} suggest that these samples are indeed dominated by major mergers. Samples with high intrinsic detected merger fractions, but no control sample \citep{canalizo_quasi-stellar_2001,urrutia_evidence_2008,glikman_major_2015,bennert_evidence_2008,veilleux_deep_2006}, are consistent with either high merger fractions in the control sample (e.g. due to mass) or a dominance of merger triggering. The possible impact of selection effects will be discussed in Section \ref{S:discussion_singlepop}.  

The enhancement of AGN fractions in mergers compared to control $r_{\textrm{merger, AGN}}/ r_{\textrm{merger, Control}}$ (Equations \ref{eqn:mergAGN},\ref{eqn:ctrAGN}) is an additional method to analyze the impact of mergers on AGN activity. Measuring this enhancement shows the duty cycle of AGN in mergers compared to control, rather than the contribution of mergers to the full AGN population. The enhancement factor depends on the merger stage and AGN type, and typically has been found to be $\sim2-7$ \citep{ellison_galaxy_2011,goulding_galaxy_2018} in the early stage of a merger, almost doubling for post mergers \citep{ellison_galaxy_2013}. The enhancement of AGN activity is also found to be increased (enhancement factor 10-20) for IR selected AGN \citep{satyapal_galaxy_2014,ellison_definitive_2019} while no merger excess is found in low emission radio galaxies once controls are matched in D4000 \citep{ellison_galaxy_2015}. 

One possible explanation of these discrepancies is that since major mergers are rare, the overall detected merger sample will have low purity even if the fraction of mergers and non-mergers identified correctly is high. The differences between mergers rates could be underestimated, as discussed by \citep{lambrides_merger_2021}. And while a highly pure sample limits the impact of this and leads to a significantly improved estimate of the merger excess in AGN, it requires much larger sample sizes and is not feasible for most studies dealing with samples sizes. In the future, machine learning models trained on simulated images will allow to calibrate false positive and false negative rates and thereby better constrain the intrinsic differences in the merger rates \citep{deepmerge,bottrell_deep_2019}, but without knowing false positive and negative rates, it is not possible to quantify the effect of this on the detected merger rates.  
\\ 
\newline

\subsection{Comparison with theoretical models}
\label{S:discussion_theory}

The connection between major mergers of galaxies and AGN has been popular in galaxy evolution models for decades,  \citet{sanders_ultraluminous_1988} suggested an evolutionary model in which AGN appear at late stages of merger after a ULIRG phase. This was later further supported by simulations \citep{di_matteo_energy_2005,hopkins_cosmological_2008}. Following this, many semi-analytical and semi-empirical models included major mergers of galaxies as the main path to black hole growth \citep{somerville_semi-analytic_2008, kauffmann_unified_2000,croton_many_2006,shankar_accretion-driven_2013}. Fig. \ref{F:mainplot} clearly disagrees with the picture that the majority of AGN are associated with major mergers. In many samples, detected merger fractions are low, and the majority of studies with control samples do not show an excess over control. Most studies that show an excess still have relatively low detected merger fractions ($\leqslant30\%$). While other studies have shown that mergers contribute to AGN fuelling, and might be prevalent in some samples, the theoretical model that all AGN are associated with major mergers of galaxies is not consistent with the data. As discussed in Section \ref{S:discussion_mergerrates}, the data so far provide upper limits only on the merger contribution to AGN, these values carry considerable uncertainties.

Other studies have suggested that the influence of major mergers depends on the luminosity. The luminosity is proportional to the accretion rate times the radiative efficiency, since the radiative efficiency only depends on the spin alignment \citep{netzer_physics_2013}, this means that over the seven order of magnitude studied here, the mass accretion rate varies by seven orders of magnitude. \citet{hopkins_characteristic_2009} suggested that there is a clear differences between low and high luminosity AGN, with a cut-off at about $L_{\textrm{bol}} \sim 10^{45}$ erg/s. From this, one would expect strong differences in the detected merger fractions at this luminosity. This is not observed (see Fig. \ref{F:mainplot}). Such a correlation is also not observed in more heterogeneous sub-samples (see Fig. \ref{F:mainplot_method}). \citet{hopkins_we_2013} analyzed the relative contribution of merger induced and stochastic fuelling, again, they find that up to redshifts $z\sim2$, which covers all studies analyzed here, merger induced fuelling dominates above $\sim L_{\textrm{bol}} 10^{46}$ erg/s (meaning the space density of merger fuelled AGN is about an order of magnitude higher than the those of stochastically fuelled AGN). Again, the data collected here do not support the picture of a transition to AGN fuelling by mergers above $\sim L_{\textrm{bol}} 10^{45-46}$ erg/s.

\citet{steinborn_cosmological_2018} studied the growth of supermassive black holes and find a mild increase in major merger fraction in high luminosities. However, the merger fraction in their study is found to depend on the galaxy mass. Therefore, the increase in detected merger fraction is due to host mass rather than luminosity, highlighting the importance of control sample. Since galaxy masses are not available across the sample, we cannot directly compare to \citet{steinborn_cosmological_2018}.

Other theoretical models emphasize the importance of other AGN fuelling mechanisms, such as disk instabilities \citep[e.g.][]{bournaud_black_2011,gabor_simulations_2013}, bars \citep{shlosman_bars_1989} or accretion from a hot halo \citep{bower_dark_2017,mcalpine_link_2017}. Disk instabilities at least at high redshift, could be identified as mergers in visual classification or morphological analysis, however, this will strongly depend on the classification used, so cannot be analysed using this compilation. The collected data clearly supports that while mergers are responsible for some AGN activity, alternative fuelling mechanisms play a substantial role in the AGN population.

\subsection{Is there a single AGN population?}
\label{S:discussion_singlepop}

While a detailed treatment of these selection effects is beyond the scope of this paper, Fig. \ref{F:mainplot_method} clearly shows that different selection methods yield  different detected merger fractions. Such a difference could either indicate that different AGN triggering mechanisms result in different observed AGN properties, with radio-loud and heavily reddened AGN preferentially found in major mergers. This would agree with other work that finds higher incidence of mergers in IR AGN \citep{goulding_galaxy_2018, ellison_definitive_2019}, higher incidence of IR compared to X-ray AGN in mergers \citep{secrest_x-ray_2020} as well as an increased level of obscuration in merging galaxies \citep[e.g.][]{ricci_growing_2017}. The higher detected merger fractions could however also be explained by selection effects, specifically, a) differences in observed images leading to differences in detectability of merger features and b) the effect of galaxy type on AGN classifications.

The detectability of merger features depends on the contrast between the central point source and the galaxy \citep[see e.g.][]{villforth_host_2019}. This means that optically bright AGN might have lower detected merger fractions simply because of the lower contamination from the point source. This could explain some of the excesses seen in radio galaxies \citep[e.g.][]{ramos_almeida_are_2011} as well as red quasars \citep[e.g.][]{urrutia_evidence_2008}. Such AGN have weaker central emission in the optical/IR wavebands in which host galaxy morphologies are studies. Based on this, one would naively expect a drop of detected merger fraction with luminosity \citep[seen for example in the samples from][]{georgakakis_host_2009}. While unfavourable contrast in optically luminous AGN could cause merger features to be washed out in optically luminous sources, there is no strong evidence for this in the data. Additionally, simulations have shown that this is not the cause of differences between high luminosity X-ray and red AGN \citep{villforth_host_2019}. While this may contribute to the higher detected merger fractions in red and radio AGN, more detailed work would be needed to determine the effect of point sources on merger detectability.

Another selection effect is the effect of the host galaxy on the AGN type. Host galaxy scale dust obscuration can lead to stronger obscuration \citep[e.g.][]{glikman_major_2015,urrutia_evidence_2008}. Since galaxies with higher dust content are also more likely to be involved in major mergers \citep{kartaltepe_candels_2014}, the merger environment could affect the AGN SED, biasing AGN types in dusty merging galaxies. Such a selection effect could explain the fact that only radio, red and IR selected sources are found to have very high detected merger fractions. However, the data collected here does not allow to test this scenario.

While selection effects can not be ruled out as a cause for the high detected merger fractions in radio and reddened AGN, the data are consistent with an evolutionary model in which AGN obscuration ages declines after a major merger, with young AGN most closely associated with mergers \citep{sanders_ultraluminous_1988, di_matteo_energy_2005, hopkins_cosmological_2008, alexander_what_2012}.

Red quasars as well as FeLoBAL quasars have been suggested to be such early or transition objects. Indeed,  several red quasar samples show high detected merger fractions \citep{urrutia_evidence_2008,canalizo_quasi-stellar_2001,donley_evidence_2018,fan_most_2016}. The candidate transition population sample of FeLoBALs \citep{villforth_host_2019} as well as another red quasar sample \citep{zakamska_host_2019} on the other hand shows low detected merger fractions (see Fig. \ref{F:mainplot_method}). While they are not seen traditionally as transition objects, radio-selected AGN also have high detected merger fraction and could potentially match the proposed population of "young" AGN.

Red quasars (and potentially radio-selected AGN) could therefore constitute a transition population with higher merger fractions, consistent with previous work showing red AGN to be associated with mergers \citep[e.g.][]{ellison_definitive_2019,goulding_galaxy_2018,secrest_x-ray_2020,kocevski_are_2015,donley_evidence_2018} and contradicting work that shows pure radio AGN to show no association with mergers \citep{ellison_galaxy_2015}. However, as discussed above, selection effects and host galaxy obscuration could also contribute to the differences seen. To distinguish between the selection effects outlined above and a transition population, AGN that are in the adult stages of current "young" red AGN would need to be identified. These "old" AGN would need to match in galaxy masses, space densities and have weaker merger features consistent with a later merger stage. Such a comparison could also be used to distinguish between the "transition" and "population" scenario discussed above.

\section{Conclusions}
\label{S:conclusion}

Here, I have presented a complete catalogue of detected merger fractions in AGN host galaxies from the literature. This catalogue contains data from 33 different studies and 50 samples. For each sample, I report consistently calculated bolometric luminosities, redshifts, detected merger fractions with consistent uncertainties as well as the AGN selection method. The AGN span a wide range in redshift ($0.025 \leq z \leq 2.0$) and luminosity ($41.5 \leq \textrm{log}(L_{\textrm{bol}} \textrm{[erg/s]}) \leq 48.0$). Detected merger fractions range from 0-100\%. 19 of the 50 samples have a control sample, of those 6 (32\%) show an excess in the merger fraction at $>$3$\sigma$ significance.  This is not an unbiased sample of AGN. The findings can be summarized as follows:


\begin{itemize}
\item In contradiction to previous work combining a smaller set of data, there is no correlation between the detected merger fraction and luminosity (Fig. \ref{F:mainplot}). This clearly contradicts theoretical models that state that major mergers dominate the AGN population at high luminosities. The lack of correlation does not appear to be due to systematic errors in the measurement of detected merger fractions. The detected merger fractions does not correlate with redshift (Fig. \ref{F:mainplot_z}).
\item Detected merger fractions in X-ray selected AGN are consistent with the uncertainties in measurements. In all but one sample, no excess in detected merger fractions over control is detected.  This indicates that the detected merger fractions in X-ray selected AGN are consistent and not strongly affected by systematic uncertainties. The contribution of major mergers to the X-ray selected AGN population is small.
\item The range in detected merger fractions for optically selected AGN is wider than for X-ray selected AGN. An excess over control merger rates is detected in one sample of low redshift high Eddington ratio AGN. The differences between samples could be due to differences in data quality or difference in the spectral energy distribution, with AGN with stronger IR excess showing higher detected merger fractions.
\item Type 2 AGN show  low detected merger rates, with excess over control seen only for very large sample sizes, indicating that while major mergers contribute to triggering in these AGN, this triggering mechanism is not dominant.
\item High detected merger fractions ($\geqslant 60\%$) are found in radio selected AGN, red quasars, as well as low-redshift optical AGN with strong IR emission. Radio selected and red AGN show larger scatter in the detected merger fraction than expected from statistical uncertainties. This indicates either intrinsic differences in the host galaxy properties for these classes of AGN or systematic uncertainties in the detected merger fractions. In both radio-selected and red quasar samples, objects also detected in the optical show lower detected merger fractions. This can be explained either by differences in the detectability of merger features or a stronger association of optically undetected red and radio AGN being preferentially triggered by mergers.
\end{itemize}

Despite the limitations of interpreting detected merger fractions, the data are in clear contradiction with a model in which the majority of AGN are associated with major mergers, as well as a picture in which major mergers become dominant at high luminosities. Triggering by major mergers clearly contributes to some AGN samples, but is not found to be dominant across the population. There is tentative evidence for higher detected merger fractions in optically undetected radio and red AGN, however, observational biases cannot be ruled out. Future studies will have to quantify the systematic uncertainties in detected merger fractions outlined here. Machine learning algorithms trained on simulated data \citep[e.g.][]{cibinel_physical_2015,bottrell_deep_2019,koppula_deep_2021} may be able to overcome the limitations hampering current analysis of merger features in AGN.


\bibliographystyle{mnras}
\bibliography{mergerrates}

\appendix

\section{Detailed comments on individual studies and full dataset}
\label{A:studies}

Detailed summary of all extracted data is given below. For a summary of the general rules that were used to extract the detected merger fractions, bolometric luminosities and any other information, see \ref{S:data}.

\begin{table*}
\tiny
\begin{tabular}{lccccccccccc}
Reference & Sub-sample & f$_{\textrm{merger, agn}}$ & n$_{\textrm{obj}}$ & f$_{\textrm{merger, ctr}}$ & Excess? & z & log(L$_{bol}$ [erg/s]) & Selection & $\lambda_{\textrm{rest}}$ [$\mu$m] & Obs & Comments\\ 
\hline
Bahcall+97$^T$ (Ba) & -- &  0.15$^{+0.11}_{-0.05}$ & 20 & -- & --  & 0.19$\pm$0.05 & 45.6$^{+0.4}_{-0.6}$ & Optical & 0.5 & HST/F606W & \\
Bennert+08 (Be) & -- &  0.80$^{+ 0.08}_{- 0.26}$ & 5 & -- & --  & 0.18$\pm0.02$ & 45.3$\pm$0.2 & Optical & 0.5 & HST/F606W & radio loud and radio-quiet\\
Boehm+12 (Bo) & -- &  0$^{+0.05}_{-0}$ & 28 & 0.07$^{+0.02}_{-0.01}$ & N & 0.5-1.1 & 43.0-45.7 & X-ray & 0.3,0.5 & HST/F606,F850LP & \\
Cales+11 (Cl) & -- & 0.38$^{+0.10}_{-0.09}$ & 29 & -- & --  & 0.2-0.4 & 44.4-45.0 & PSB & 0.5 & HST/F606W & Post-star-burst\\
Canalizo+01 (Cn) & -- &  1$_{-0.1}$ & 9 & -- & --  & 0.1-0.4 & 45.8$\pm$0.4 & Red AGN & 0.5-0.7 & HST/F606W,F702W,F814W & ULIRG/QSO \\
Chiaberge+15 (Ch3c) & Hz3C &  1.0$^{+0.0}_{-0.09}$ & 11 & 0.27$^{+ 0.08}_{- 0.06}$ & Y  & 1.4$\pm$0.4 & 44-46 & Radio & $\sim$1.0 & HST/F140W & High-power 3C radio galaxies\\
Chiaberge+15 (ChHzR) & HzLLRG & 0.88$^{+0.05}_{-0.24}$ & 8 & 0.27$^{+ 0.08}_{- 0.06}$ & Y  & 1.3$\pm$0.3 & 41-43 & Radio & $\sim$1.0 & HST/IR & Low-power radio galaxies\\
Chiaberge+15 (ChLP2) & LPType2 &  0.38$^{+0.10}_{-0.08}$ & 26 & 0.20$^{+ 0.07}_{- 0.05}$ &  N  & 1.4$\pm$0.2 & 43.0$^{+0.2}_{-0.3}$ & Type 2 & $\sim$1.0 & HST/IR & Low power Type2 AGN\\
Chiaberge+15 (ChHP2) & HPType2 &  0.33$^{+0.18}_{-0.11}$ & 9 & 0.20$^{+ 0.07}_{- 0.05}$ & N & 1.8$\pm$0.3 & 45.9$^{+0.1}_{-0.2}$& Type2 & $\sim$1.0 & HST/IR & High power Type2 AGN\\
Cisternas+11$^T$ (Ci) & -- &  0.15$\pm$0.09 & 140 & 0.13$\pm$0.07 & N  & 0.3-1.0 & 43.0-45.7 & X-ray & 0.4-0.6 & HST/F814W & \\
del Moro+16 (dM) & -- & 0.30$^{+0.09}_{-0.07}$ & 33 & -- & -- & 2.2$\pm$0.5 & 45.9$\pm$0.3 & IR & $\sim$0.5 & HST/F160W & \\
Donley+18 (DoIR) & Red Quasar & 0.5$\pm$0.12 & 16 & -- & --  & 0-5 & 44-46.5 & Red AGN & $\sim$0.1-1.6 & HST/F814W,F125W,F160W & Sanders IR only \\
Donley+18 (DoX) & X-ray & 0.26$^{+0.6}_{-0.5}$  & 64 & -- & --  & 0-5 & 43-47 & X-ray & $\sim$0.1-1.6 & HST/F814W,F125W,F160W & full X-ray only\\
Dunlop+03 (DQ) & RQQ &  0.46$^{+ 0.14}_{- 0.12}$ & 13 & -- & --  & 0.1-0.25 & 45.5$\pm$0.2  & Optical & $\sim0.5$ & HST/R & radio-quiet quasars\\
Dunlop+03 (DL) & RLQ &  0.10$^{+ 0.17}_{- 0.04}$ & 10 & -- & --  & 0.1-0.25 & 45.9$\pm$0.1 & Radio & $\sim0.5$ & HST/R & radio-loud quasars\\
Dunlop+03 (DG) & RG &  0.00$^{+ 0.10}_{- 0.00}$ & 10 & -- & --  & 0.1-0.25 & 41-45 & Radio & $\sim0.5$ & HST/R & radio galaxies \\
Ellison+19 (E) & -- &  0.31$\pm0.01$ & 1269 & 0.16$\pm0.01$  & Y  & 0-0.25 & 41.9-44.6 & Type 2 & $\sim0.5$ & CFIS/r & \\
Fan+16 (F) & -- & 0.61$^{+ 0.10}_{- 0.12}$  & 18 & --  & --  & 2.8$\pm$0.7 & 47.5$\pm$0.3 & Red AGN & $\sim$0.3-0.4 & HST/F110W,F160W & HotDOG\\
Georgakakis+09$^T$ (Ge) & all & 0.34$\pm$0.15  & 266 & - & -  & 0.5-1.3 & 41.7-47.0 & X-ray & $\sim$0.2-0.7 & HST/F435W,F606W,F775W,F850LP & full sample \\
Georgakakis+09$^T$ (Ge41) & L41 &  0.41$\pm$0.22 & $\sim$70 & - & -  & 0.5-1.3 & 41.7-43.0 & X-ray & $\sim$0.2-0.7 & HST/F435W,F606W,F775W,F850LP & 41 < log($L_X$ [erg/s]) < 42\\
Georgakakis+09$^T$ (Ge42) & L42 &  0.16$\pm$0.05& $\sim$70 & - & -  & 0.5-1.3 & 43.0-44.0 & X-ray & $\sim$0.2-0.7 & HST/F435W,F606W,F775W,F850LP & 43 < log($L_X$ [erg/s]) < 43\\
Georgakakis+09$^T$ (Ge43) & L43 &  0.17$\pm$0.06 & $\sim$70 & - & -  & 0.5-1.3 & 44.0-45.7 & X-ray & $\sim$0.2-0.7 & HST/F435W,F606W,F775W,F850LP & 43 < log($L_X$ [erg/s]) < 44\\
Georgakakis+09$^T$ (Ge44) & L44 &  0.05$\pm$0.05 & $\sim$70 & - & -  & 0.5-1.3 & 45.7-47.0 & X-ray & $\sim$0.2-0.7 & HST/F435W,F606W,F775W,F850LP & 44 < log($L_X$ [erg/s]) $<$ 45\\
Glikman+15 (Gl) & -- &  0.80$^{+0.07}_{-0.17}$ & 10 & -- & --  & 2.0$\pm$0.2 & 47.8--48.3 & Red AGN & 0.5 & HST/F160W & \\
Grogin+05 (Gr) & -- &  0.08$\pm$0.02 & 322 &  0.08$\pm$0.02 & N  & 0.4-1.3 & 43-45.7 & X-ray  & 0.3-0.6 & HST/F606W,F775W,F850LP & \\
Hewlett+17 (Hlz) & low-z &  0.06$^{+0.02}_{-0.01}$ & 35 & 0.06$^{+0.4}_{-0.02}$ & N  & 0.5-1.1 & 45.0-46.4 & X-ray & 0.5 & HST/F814W & Low redshift ($\sim 0.8$)\\
Hewlett+17 (Hmz)& mid-z &  0.06$^{+0.02}_{-0.01}$ & 35 & 0.034$^{+0.01}_{-0.005}$ & N  & 1.1-1.7 & 45.0-46.4 & X-ray & 0.3 & HST/F814W & Low redshift ($\sim 0.8$)\\
Hewlett+17 (Hhz) & high-z &   0.04$^{+0.03}_{-0.01}$ & 35 & 0.008$^{+0.007}_{-0.003}$ & N  & 1.7-2.1 & 45.7-47 & X-ray & 0.3 & HST/F814W & high redshift ($\sim 1.9$)\\
Hong+15 (Ho) & -- & 0.44 $\pm$ 0.08 & 39 & - & - & 0.16$\pm$0.06 & 45.7$\pm$ 0.5 & Optical & $\sim$0.4-1 & various ground & \\
Hutchings+84 (HR)& Radio &  0.35$^{+ 0.10}_{- 0.08}$ & 26 & -- & --  & 0.0-0.7 & 45.44$\pm$0.8 & Radio & $\sim$0.3-0.6 & CFHT/B,R & \\
Hutchings+84 (HX) & X-ray &  0.29$^{+ 0.09}_{- 0.07}$ & 29 & -- & --  & 0.0-0.7 & 44.7$\pm$0.7l & X-ray & $\sim$0.3-0.6 & CFHT/B,R &  \\
Hutchings+84 (HO) & Optical & 0.17$^{+ 0.11}_{- 0.05}$ & 23 & -- & --  & 0.0-0.7 & 44.9$\pm$0.7 & Optical & $\sim$0.3-0.6 & CFHT/B,R &  \\
Kocevski+12$^T$ (K12) & -- &  0.17$^{+0.05}_{-0.04}$ & 72 & 0.18$^{+0.03}_{-0.02}$ & N  & 1.5-2.5 & 43.0-45.7 & X-ray & 0.5 & HST/F160W & \\
Kocevski+15 (K15lN) & low N$_H$ &  0.21$^{+0.03}_{-0.02}$ & ~100 & -- & --  & 0-2 & 43.0-45.5 & X-ray & $\sim$0.5-1.5 & HST/WFC3 & \\
Kocevski+15 (K15mN) & moderate N$_H$ &  0.31$^{+0.03}_{-0.03}$  & ~100 & -- & --  & 0-2 & 43.0-45.5 & X-ray & $\sim$0.5-1.5 & HST/WFC3 & \\
Kocevski+15 (K15hN) & high N$_H$ &  0.43$^{+0.05}_{-0.04}$  & $\sim$100 & -- & --  & 0-2 & 43.0-45.5 & X-ray & $\sim$0.5-1.5 & HST/WFC3 & \\
Koss+10$^T$ (Ko) & -- &  0.18$^{0.05}_{-0.04}$ & 72 & 0.01 & Y  & 0.0-0.05 & 44.8$\pm$0.6 & X-ray & 0.3-0.8 &SDSS+KittPeak/ugriz & \\
Liu+09  (L) & -- &  0.22$^{+0.18}_{-0.08}$ & 9 & -- & --  & 0.4-0.6 & 46$\pm$0.2 & Type 2 & 0.2-0.5 & SDSS/ugirz? & \\
Marian+19 (Ma19) & -- &  0.24 $\pm$ 0.09 & 21 & 0.19 $\pm$ 0.04 & N  & 1.9-2.1 & 46.8$\pm$0.2 & Optical & $\sim$0.5 & HST/F160W & \\
Marian+20 (Ma20) & -- &  0.14$\pm$0.12 & 17 & 0.08$\pm$0.06 & Y  & 0.15$\pm$0.03 & 46.3$\pm$0.3 & Optical & $\sim$0.4,0.5 & ESO VLT/FORS2 B, V & \\
Mechtley+16 (Me) & -- &  0.39$\pm$0.11 & 19 & 0.30$\pm$0.05 & N  & 1.9-2.1 & 47.3$\pm$0.3 & Optical & $\sim$0.5 & HST/F160W &  log(M$_{BH}$ [$M_{\odot}$])$>9$\\
Ramos-Almeida+11 (RA) & -- &  0.88$^{+0.02}_{=0.06}$ & 46 & 0.49$\pm$0.04 & Y  & 0.05-0.7 & 40.0-45.0 & Radio & ? (optical) & Gemini/? & \\
Urrutia+08$^T$ (U) & -- &  0.85$^{+0.05}_{-0.14}$ & 13 & -- & --  & 0.4-1.0 & 45.7-46.7 & Red AGN & $\sim$0.3, 0.5 & HST/F475W,F814W & \\
Veilleux+09 (Ve) & -- &  0.57$^{+ 0.09}_{- 0.10}$ & 28 & -- & --  & 0.14$\pm$ 0.08&  45.7$\pm$0.5 & Optical & $\sim$1.5 & HST/H & \\
Villforth+14  (V14) & -- & 0.07$^{+ 0.05}_{- 0.02}$  & 60 & 0.07$^{+ 0.01}_{- 0.01}$ & N  & 0.5-0.8 & 41.7--46.4 & X-ray & 1.0 & HST/F160W & \\
Villforth+17 (V17) & -- &  0.20$^{+ 0.12}_{- 0.06}$ & 20 & 0.22$^{+ 0.05}_{- 0.04}$ & N  & 0.5-0.7 & 47.3$\pm$0.4 & Optical & $\sim$1.0 & HST/F160W &  \\
Villforth+19 (V19) & - &  0.30$^{+ 0.17}_{- 0.10}$ & 10 & -- & --  & 0.6-1.0 & 46.7$\pm$0.4 & Red AGN & $\sim$1.0 & HST/F160W & FeLoBAL\\
Wylazalek+16 (W)& -- & 0.30$^{+ 0.12}_{- 0.08}$ & 20 & -- & -- & 0.5$\pm$0.1 & 45.6$\pm$0.3 & Type 2 & $\sim$0.3/0.5 & HST & \\
Zakamska+19 (Z2) & Type 2 &  0.00$^{+ 0.15}_{- 0.00}$ & 6 & -- & --  & 2-3 & 45-47 & Type 2 & $\sim$0.2-0.5 & HST/F814W,F160W &\\
Zakamska+19 (ZR) & Red Quasars &  0.10$^{+ 0.17}_{- 0.04}$ & 10 & -- & --  & 2-3 & 46-48 & Red AGN & $\sim$0.2-0.5 & HST/F814W,F160W &\\
\end{tabular}
\caption{Compilation of merger rates  from the literature, in alphabetical order. Reference: reference for merger fractions, abbreviation used for figures is given in brackets, $^T$ indicates studies includes in \citet{treister_major_2012}; Sub-sample: description of sub-sample (for studies with several samples); f$_{\textrm{merger, agn}}$: detected merger fraction; n$_{\textrm{obj}}$: sample size;  f$_{\textrm{merger, control}}$: fraction of mergers in control sample, if applicable; Excess: flag indicating excess of merger rate above control, --: not applicable, 'N': no excess, 'Y': excess detected; z: redshift; log(L$_{\textrm{bol}}$ [erg/s]): bolometric luminosity, for detail, see Appendix \ref{A:studies}; Selection: selection method for the sample, for detail see Section \ref{S:data_selection} and Appendix \ref{A:studies};  $\lambda_{\textrm{rest}}$ [$\mu$m]: approximate rest wavelength; Obs: Telescope/Filter for observations; Comments: comments and notes.}
\label{maintable}
\end{table*}

\subsection{Bahcall+97 (Ba)}

\citet{bahcall_hubble_1997} used a combination of archival data and targeted observations to study 20 z$<$0.3 AGN selected from the Veron-Cetty Quasar Catalogue. For the merger fraction, I cite the fraction of objects cited to be in gravitationally interacting systems in Table 2 not including those with non-interacting companions (3/20). The average and median absolute magnitude are $M_V$ -23.4/-23.2. The magnitude system is not given, but since Vega and AB magnitudes are near identical in the V band, we calculate the luminosities assuming AB, this is not expected to introduce errors. I convert these absolute magnitudes to erg/s using 5100\AA as the wavelength and then convert to bolometric luminosities following \citep{netzer_physics_2013}.

\subsection{Bennert+08 (Be)}

\citet{bennert_evidence_2008} studied 5 radio-loud and radio-quiet AGN at z$\sim$1.8 also analyzed by \citet{dunlop_quasars_2003} with considerably deeper HST imaging in F606W. The sample is optically selected. 4/5 objects show signatures of mergers, this is used as the detected merger fraction. For the bolometric luminosities, I use the 2D decomposition to derive absolute magnitudes in F606W and convert them to bolometric luminosities using the $\nu L_{\nu}$ relations from \citep{bennert_evidence_2008}. \citet{bennert_evidence_2008} perform several different 2D fits for a PSF, PSF+disk, PSF+bulge and PSF+disk+bulge. I use the fit with the best reduced $\chi^2$, which is in all cases the PSF+disk+bulge fit.

\subsection{Boehm+12 (Bo)}

\citet{boehm_agn_2012} studied intermediate luminosity ($L_X < 10^{44}$ erg/s) at z$\sim$0.7 in the GEMS and Stages surveys using HST F606W. They perform both visual classification and quantitative morphological analysis, both show no excess of merger fractions in the AGN. For the merger fractions, \citet{boehm_agn_2012} use eight human classifiers and cite merger fraction for which 4,5,6 or 7 or eight classifiers agree, the four classifier results are used, giving the highest merger fractions. X-ray luminosities are used to derive bolometric luminosities.

\subsection{Cales+11 (Cl)}

\citet{cales_hubble_2011} studied the host galaxies of 29 post-star-burst quasars at z$\sim$0.3 using HST in F606W. For the detected merger fraction, I use the disturbance flag in Table 2  (11/29). For the bolometric luminosity derivation, I use the absolute magnitudes of the PSF component in F606W ($\lambda_{\textrm{rest}}\sim 4500\AA$) given in Table 3 and 4. These PSF magnitudes are derived from 2D fits of a host galaxy and point sources to the HST imaging. The bolometric luminosities given are for the mean and standard deviation.  The magnitudes are converted to bolometric luminosities given the relation for the 5100\AA magnitude following \citet{netzer_physics_2013}.

\subsection{Canalizo+01 (Cn)}

\citet{canalizo_quasi-stellar_2001} selected 9 ULIRG/QSO mixture objects at low redshift z$\sim$0.1-0.4. Since these AGN are selected in the IR to have colours in between ULIRGs and AGN, we classify these as IR selected /red AGN. All AGN in their sample show signs of mergers. For the bolometric luminosities, I use the B-band magnitudes without a k-correction in combination with the bolometric corrections from \citet{netzer_physics_2013}.

\subsection{Chiaberge+15 (Ch3c, ChHzR, ChLP2, ChHP2)}

\citet{chiaberge_radio_2015} studied 4 different samples of Type 2, radio-loud, and radio-quiet AGN in HST/IR at redshift z$\sim$1-2. Two sets of control galaxies are compared in visual classification to the radio-loud and radio-quiet AGN. We calculate the excess probabilities by considering beta distributions following \citep{cameron_estimation_2011} and find excesses in the radio-loud samples, a marginal excess in one radio-quiet sample, and no excess in the other.

The samples are as follows, bolometric luminosities are derived as follows:

The radio-loud samples are described as follows. Both are listed as Radio selected.
\begin{itemize}
\item High redshift 3C galaxies (Hz3C, data in Table 1 in \citet{chiaberge_radio_2015}).The redshift range is 1.1-2.5. The radio power has a narrow range of $\textrm{log P [erg/s/Hz]} ~ 35.4 \pm 0.3$.
\item High redshift low luminosity radio galaxies (HzLLRG, data in Table 2 in \citet{chiaberge_radio_2015}). The redshift range is slightly lower (1-1.6). Radio powers are considerable lower at $\textrm{log P [erg/s/Hz]} ~ 31.2 \pm 0.6$.
\end{itemize}

For both, we follow the approach used for radio studies, as outlined in \ref{S:bolcor}. The luminosities for the 3C sample are slightly outside the range studied in \citet{best_fundamental_2012}, I extrapolate linearly. There is a rather large uncertainty in the bolometric luminosities, reflected by the error bars given. Both radio-loud samples show an excess of merger detections to the matched control sample of bright galaxies. The radio-loud samples are treated as radio selected.

Additionally, \citep{chiaberge_radio_2015} present two samples of radio-quiet Type 2 AGN, a bright sample with $1 < z < 2.5$ and $\textrm{log}(L_X \textrm{[erg/s]}) > 44$ (ChHP2, Table 4 in \citet{chiaberge_radio_2015}) and a low luminosity sample with the same redshift range and $\textrm{log}(L_X \textrm{[erg/s]}) < 42$ (ChLP2, Table 3 in \citet{chiaberge_radio_2015}). I list these as Type 2, rather than X-ray selected. I apply bolometric corrections to the X-ray luminosities following \citep{netzer_physics_2013}.

\subsection{Cisternas+11 (Ci)}

\citet{cisternas_bulk_2011} study 140 X-ray selected AGN at redshift z=0.3-1 in the COSMOS field in HST/F814W. They use visual classification and report consolidated scores from 10 classifiers. They use a control sample. \cite{cisternas_bulk_2011} report two disturbance classes, 1 and 2, where 1 is reporting mild disturbances and 2 major disturbances, I use class 2. For bolometric corrections, I use the conversion from \citep{netzer_physics_2013} from the X-ray luminosities.

\subsection{del Moro+16 (dM+16)}

\citet{del_moro_mir_2016} studied mid-IR selected AGN. For the bolometric luminosities, I use the SED fitting based luminosities given in Table 1. For the detected merger rates, \citet{del_moro_mir_2016} follow the visual classification used by \citep{kocevski_are_2015}. I use the disturbed classification and combine both the obscured and unobscured sample. using the datapoints given in Figure 6. Error bars are recalculated following \citet{cameron_estimation_2011}.

\subsection{Donley+18 (DoIR, DoX)}

\citet{donley_evidence_2018} study X-ray and IR selected AGN in COSMOS. They analyse multi-band data (F814W, F125W, F160W). The redshift range is $0<z<5$. There are several different samples, mostly divided in X-ray and IR selected. For simplicity, we select the largest X-ray only sample (64 AGN) as well as the largest IR only sample (Sanders sample, 16 AGN). 

The detected merger fractions here are given in several different categories, I inspect the snapshots with classifications in their Fig.4 to determine which interaction class to use. Since I am looking for significant disturbances, the Interacting/Merging most closely matches this. I do not include the disturbed category since these galaxies show on average very weak disturbances (see their Table 1).
 
For the bolometric luminosities, \citep{donley_evidence_2018}  use SED as well as X-ray spectral fitting. Since no raw luminosities are given, the bolomtric luminosities from SED fitting are used. No full table is given, so we extract the range of luminosities from Figure 2, left panel.

\subsection{Dunlop+03 (DQ, DL, DG)}

\citet{dunlop_quasars_2003} studied 13 radio-quiet quasars (RQQs), 10 radio-loud quasars (RLQs), and 10 radio galaxies (RGs) at $0.1<z<0.25$ using HST WFC2 R band data. The RLQ and RLQ samples are listed as radio selected and the RQQ sample as optically selected. For the detected merger fractions, I use the visual classifications in Section 7.4, Table 8. I use the 'tidal' classification for the detected merger fraction. \citet{dunlop_quasars_2003} perform a comparison to cluster galaxies, but since they are not matched specifically to the hosts, I treat this sample as having no control. No excess over control would have been detected if we had included the control.

For bolometric luminosities for the RQQ and RLQ sample, I use the nuclear luminosities in R cited in Section 6.22. The R band rest closely matches $5100\AA$, so we use the bolometric correction for this wavelength following \citep{netzer_physics_2013}.

For the radio galaxies, I derive the bolometric luminosities following the approach outlined in Section \ref{S:bolcor}. 

\subsection{Ellison+19 (E)}

\citet{ellison_definitive_2019} analyse CFIS r band analysis of 1124 Type 2 AGN as well as 245 IR selected AGN at $0< z < 0.25$. AGN are visually inspected and compared to a matched control sample. For the detected merger fraction, I use the classification of both post-mergers and interacting pairs. An excess of the rate of mergers in AGN over control is detected. The inclusion of interacting pairs does not affect the results regarding excess over control.

\citet{ellison_definitive_2019} do not provide bolometric luminosities for their objects. For the Type 2 AGN, [OIII] luminosities are given. These are used to calculate bolometric luminosities following \citet{lamastra_bolometric_2009}. Since the distribution of luminosities is not given, I use the range of [OIII] luminosities as $40 < \textrm{log}(L_{\textrm{[OIII]}} \textrm{[erg/s]}) < 42$. The cited luminosities therefore represent the luminosity range of the sample. No information is available on the luminosities of the IR selected AGN, they are therefore not included in the analysis.

\subsection{Fan+16 (F)}

\citet{fan_most_2016} study 18 HotDOGs with AGN activity using HST WFC3 imaging. The sample is listed as red/IR selected. For the bolometric luminosities, they use decomposition of the IR SED, this is used here for the bolometric luminosity. 

\subsection{Georgekakis+09 (Ge, Ge41, Ge42, Ge43, Ge44)}

\citet{georgakakis_host_2009} studied AGN host galaxies in the GOODS fields in HST bands F435W, F606W, F775W, and z/F850LP over a redshift range of 0.3 $<z<$1.3. Visual inspection is performed and objects are classified as disk, early types, peculiar and point sources. I use the merger fractions reported in Table 3. The X-ray luminosity range is log($L_X$)41-46, subdivided into samples 41-42, 42-43, 43-44, 44-45, we discard the sample with luminosities 45-46 since it is point-source dominated. Bolometric luminosities are calculated based on the X-ray luminosities following \citet{netzer_physics_2013}.

\subsection{Glikman+15 (Gl)}

\citet{glikman_major_2015} studied 10 red quasars at z$\sim$2 using HST F160W. Due to the unusual SEDs of the samples, the bolometric luminosities given in \citet{glikman_major_2015} are used directly. Point source magnitudes are given, but since those are heavily affected by dust, they are likely to underestimate the luminosity. For the detected merger fractions, the value 8/10 cited in the paper is used, although this includes a large number of sources with close neighbours.

\subsection{Grogin+05 (Gr)}

\citet{grogin_agn_2005} did not report merger fractions, but instead compared the asymmetry \citep[see e.g.][]{conselice_asymmetry_2000,conselice_fundamental_2006}, to match \citet{villforth_morphologies_2014}, who used asymmetry for comparable data, I use a cut-off of A=0.1 to derive the merger fractions. Luminosities are not directly given but are cited to be $\textrm{log}(L_X \textrm{[erg/s]})>42$. Since the sample has large overlap with other samples studied in the same field \citep{boehm_agn_2012,hewlett_redshift_2017,villforth_morphologies_2014}, I assume a luminosity range $\textrm{log}(L_X \textrm{[erg/s]})\sim42-44$ and apply bolometric corrections to the X-ray luminosities following Section \ref{S:bolcor}.

\subsection{Hewlett+17 (Hlz, Hmz, Hhz)}

\citet{hewlett_redshift_2017} studied X-ray selected AGN in the COSMOS fields. For the detected merger fraction, I use the merger classification given. Bolometric luminosities are calculated from X-ray luminosities following Section \ref{S:bolcor}.

\subsection{Hong+15 (Ho)}

\citet{hong_correlation_2015} studied Type 1 selected AGN using a number of ground based telescoped. The sample is listed as optically selected. Bolometric luminosities are calculated using the nuclear absolute magnitudes from their Table 1.

\subsection{Hutchings+84 (HR, HX, HO)}

\citet{hutchings_optical_1984} studied the host galaxies of 78, divided into radio, X-ray and optical AGN. The redshift range is 0.0-0.7. No control sample is used.

For the visual classification, three categories are given in Table 2, "interacting" (5-10\%), probable interacting (17-35\%), possible interacting (17-46\%), and interacting or close companion (30-54\%). No images are available. From the description in the text, "interacting" are clear train-wreck mergers, while "probable interacting" still shows strong signs of interaction. Therefore, the "probable interacting" category is used. For the fractions, I calculate the number of sources given the fractions cited and assuming integer values, we then recalculate errors following\citep{cameron_estimation_2011}.

For the bolometric luminosity, Table 1 lists properties of the full sample. All objects have nuclear luminosities in R available, B magnitudes are only available for a small subset. X-ray and radio luminosities are available for a sub-sample. However, even some of the X-ray and Radio objects are missing luminosities in their respective bands. Nuclear luminosities are therefore used in all cases for consistences. Since the redshifts are low, the R band is close to the 5100\AA, bolometric corrections based on $\nu L_{\nu}$ from \citep{netzer_physics_2013} are used.

\subsection{Kartaltepe+10 (not used)}

\citet{kartaltepe_multiwavelength_2010} studied Luminous Infra-red Galaxies (LIRGs), Ultra-LIRGs (ULIRGs) and Hyper-LIRGs (HyLIRGs) in COSMOS. This study is one of those included in \citep{treister_major_2012}. \citet{kartaltepe_multiwavelength_2010} derive the fraction of objects hosting AGN, but not the AGN luminosity. Since no AGN luminosities are available, this study is not used.

\subsection{Kocevski+12 (K12)}

\citet{kocevski_candels:_2012} used a visual classification following \citet{kartaltepe_candels_2014} to study the hosts of 72 X-ray selected AGN at redshift z=1.5 to 2.5 using HST F160W. AGN hosts show no excess of disturbances compared to control. For detected merger fractions, the 'Disturbed' category is used, selecting only galaxies with strong signs of interaction. The luminosities cited are derived using the X-ray luminosities given and bolometric correction following \citet{netzer_physics_2013}.

\subsection{Kocevski+15 (K15lN, K15mN, K15hN)}

\citet{kocevski_are_2015} analyzed HST WFC3 F160W imaging of X-ray selected AGN as a function of obscuration and performed visual classification. The three sub-samples are included separately.

For the detected merger fraction, I use the disturbed fraction and do not reject point sources from the sample. Note that the point source fractions were lower in the higher obscuration sample. For the bolometric luminosities, I give the value for the full range of X-ray luminosities and correct using \citep{netzer_physics_2013}.

\subsection{Koss+10 (Ko)}

\citet{koss_merging_2010} studied the incidence of mergers in Hard X-ray selected BAT AGN at $z<0.05$. The detected merger rate is determined using visual inspection and compared to a control sample of inactive galaxies. For the detected merger fraction, the fraction of disturbed morphologies is used (18\% in AGN compared to 1\% in control). \citet{koss_merging_2010} also perform a comparison to Type 2 AGN, however, this is only done for the incidence of neighbours, rather than disturbances, so this is not used. \citet{koss_merging_2010} do not give luminosities in their paper. Since the parameter space in this study is unique and not covered by other studies included, I estimate the luminosities using SWIFT BAT catalogue \citep{oh_105_2018}. I select all sources with a redshift $z<0.05$, this yields 527, I assume that the sample used in \citet{koss_merging_2010} is a random sub-sample of this sample. The matched AGN have an average luminosity log(L$_{BAT}$ [erg/s]) = 43.3 $\pm$ 0.6. Since I could not find reliable bolometric corrections for the hard X-ray and do not have information about the spectral shape across the sample, I use the soft X-ray bolometric corrections following \citep{netzer_physics_2013}.

\subsection{Lanzuisi+15 (not used)}

\citet{lanzuisi_compton_2015} performed a detailed study of Compton-thick AGN in COSMOS. The 10 Compton-thick AGN in their work span a very wide range of redshifts (0.125-2.429) and report bolometric luminosities for 9/10 sources and morphological classification for 8/10. Since it is not clear from the paper which subset of their parent sample was used for morphological analysis, this study is not included in the table, but discussed in the context of high detected merger fractions in obscured samples.

\subsection{Liu+09 (L)}

\citep{liu_host_2009} studied 9 luminous Type 2 AGN. For the detected merger fraction, the number of sources that have double cores (2/9) is used, sources with close neighbours are not included. For the bolometric luminosities, I use the [OIII] luminosities given on the paper and apply corrections following Section \ref{S:bolcor}.

\subsection{Marian+19 (Ma19)}

\citep{marian_major_2019} selected high Eddington ratio (L/L$_{edd}>$0.7) AGN from SDSS. For bolometric luminosities, the SDSS derived bolometric luminosities from Table 1 derived using optical luminosities, this closely matches the bolometric correction applied in other cases following \citet{netzer_physics_2013}.

Merger fractions are determined as follows: human classifiers ranked both AGN (n=21) and control galaxies (n=192) by merger status. The order determined by classifiers was used to determine if there is a statistically significant difference between AGN and control. Merger fractions cited in the table were determined by choosing a cut-off for detectable mergers.

\subsection{Marian+20 (Ma20)}

\citep{marian_significant_2020} selected high-Eddington ratio optically unobscured AGN ($L/L_{edd} > 0.3$) from the Hamburg/ESO survey, the Palomar Green Survey (PG), and the
SDSS DR7. Bolometric luminosities are calculated from the optical luminosities listed in Table 1 following Equation \ref{EQ:lbol_5100}. The merger fractions are determined using the same method as given in \citep{marian_major_2019}.  

\subsection{Mechtley+16 (Me)}

Objects in \citet{mechtley_most_2016} are selected from SDSS to have high black hole masses, rather than high luminosities. We use the k-corrected V band magnitudes given in Table 1 and apply bolometric corrections following \citet{netzer_physics_2013}. 

Merger fractions are determined as follows: human classifiers ranked both AGN (n=19) and control galaxies (n=160) by merger status, the order determined by classifiers was used to determine if there is a statistically significant difference between AGN and control. Merger fractions cited in the table were determined by choosing a cut-off for detectable mergers.

\subsection{Ramos-Almeida+11 (RA)}

\citet{ramos_almeida_are_2011} studied the hosts of radio galaxies at low redshift, compared to a matched control sample. The detected merger fractions as cited in the paper are used.

\citet{ramos_almeida_are_2011} do not give luminosities, however, the sample is the same as in \citet{tadhunter_optical_1993}, who cite [OIII] luminosities. Using the range of [OIII] luminosities from \citet{tadhunter_optical_1993}, bolometric corrections are applied following following Section \ref{S:bolcor}. The large uncertainty in this correction is reflected by the error bar.

\subsection{Schawinski+11 (not used)}

\citet{schawinski_hst_2011} studied 23 X-ray selected AGN in HST HF160W from Early Release Science Data. They did not report detected merger fractions, so this dataset is not used. With X-ray luminosities $42 < \textrm{log}(L_X [erg/s]) < 44$ and redshifts $1.5 < z < 3$, the parameter space covered in this study overlaps with that of \citet{kocevski_candels:_2012}. \citet{schawinski_hst_2011} find that the general morphological properties agree with a matched control sample, in agreement with \citet{kocevski_candels:_2012}. Therefore, omission of this study will not bias our findings.

\subsection{Schawinski+12 (not used)}

\citep{schawinski_heavily_2012} studied the host galaxies of HotDOGs using HST/WFC3. They found the hosts to be disk dominated with low detected merger fraction. This study is not used because bolometric luminosities cannot be derived. With a redshift of $z\sim2$, this part of parameter space is covered by several study finding both high detected merger fractions and no excess compared to control \citep{kocevski_candels:_2012, hewlett_redshift_2017}. 

\subsection{Urrutia+08 (U)}

\citet{urrutia_evidence_2008} studied 13 extremely red quasars using HST in two bands (F475W, F814W). As a detected merger fraction, we use the stated value of 11/13 showing strong signs of mergers. This sample is listed as a Red/IR selected.

The abstract states that the dust-corrected B-band magnitudes are -23.5 $< M_B < $-26.2, bolometric luminosity corrections are applied to these values following Section \ref{S:bolcor}. 

\subsection{Veilleux+09 (Ve)}

\citep{veilleux_deep_2009} studied 28 low-redshift quasars using HST/NICMOS. Since they are PG selected quasars, they are listed optically selected. They report 16/28 to show merger signatures, which they describe as "Signs of galactic interactions such as tidal tails and bridges, lopsided disks, distorted outer isophotes, or double nuclei."  This is used as the detected merger fraction.

For the bolometric luminosity, two choices are available: the IR reported PSF magnitudes from the image decomposition or the bolometric luminosities reported in the paper. The PSF magnitudes are given in the IR, and so no reliable bolometric corrections are available. The bolometric luminosities are based on L(5100\AA), but also use the contribution from the IR. Given the uncertainty of the PSF fits as well as the fact that the bolometric corrections reported are closer to our general methodology, the bolometric luminosities from \citet{veilleux_deep_2009} are used.

\subsection{Villforth+14 (V14)}

\citet{villforth_morphologies_2014} use X-ray selected AGN in GOODS-S in the redshift range 0.5-0.8 with X-ray luminosities in the range $41 < L_X \textrm{[erg/s]} < 45.5$. We use bolometric corrections from \citep{netzer_physics_2013} to derive bolometric luminosities from the X-ray luminosities. 

\citep{villforth_morphologies_2014} used visual classifications mostly for cross-checking and concentrated on the distribution of the measured asymmetries instead. Specifically, we use the cut-off of A=0.1 as used in the paper to determine the detected merger fraction. 

\subsection{Villforth+17 (V17)}

\citet{villforth_host_2017} selected SDSS quasars based on their X-ray luminosities to have log(L$_{log}$ [erg/s]) $>$45. They use a mixture of quantitative morphology measures (asymmetry and shape asymmetry, see \citep{pawlik_shape_2016}) as well as visual classification. Both are compared to a control sample of simulated AGN and show no excess of merger fractions in the AGN sample. Merger fractions from visual inspection are used, specifically, we use the "disturbed" rates, which include both clear mergers and weaker disturbances, although the two numbers are comparable. To make a comparison with other datasets straightforward, the detected merger rates without rejecting unresolved sources to make data more comparable to the wide variety of datasets not removing point sources from the sample \citep[e.g.][]{georgakakis_host_2009,hewlett_redshift_2017}.

For consistency with other studies cited here, I use the X-ray luminosities from \citep{villforth_host_2017} and correct them using X-ray bolometric corrections following Section \ref{S:bolcor}.

\subsection{Villforth+19 (V19)}

\citet{villforth_host_2019} studied the host galaxies of 10 FeLoBALs and compare them to the blue quasars from \citet{villforth_host_2017}. The objects were observed with HST WFC3 in the F160W/H band. \citep{villforth_host_2019} find no excess compared to the control quasars from \citet{villforth_host_2017} and \citet{villforth_host_2017} find no excess over matched control galaxies. Since no direct comparison to matched galaxies is performed, I list this study as having no control sample. Following \citet{villforth_host_2017}, the visual classification for disturbed (3/10) rather than merger (1/10) is used. 

Determining bolometric corrections for FeLoBALs can be challenging due to the high levels of reddening and X-ray weakness \citep{morabito_suzaku_2011, dai_intrinsic_2012,dunn_determining_2015}. We therefore use the fitted quasar magnitude in the H band (rest frame $\sim 1 \mu m$). We then use the relations from \citep{netzer_physics_2013} for $\nu L_{\nu} (5100\AA)$. Due to the uncertainty about the SED of FeLoBALS, we do not apply a k-correction to transfer between $\sim10000\AA$ and   $\sim5000\AA$. If we assume that the FeLoBAL SEDs are intrinsically similar to normal quasar SEDs, this would mean I underestimate the bolometric luminosity by $\sim$ 0.5 dex.

\subsection{Wylezalek+16 (W)}

\citet{wylezalek_towards_2016} analyzed 20 Type 2 AGN. The study lists two samples, but since the selection is very similar and only the origin of the data is different, both are combined here. The bolometric luminosities are calculated from the [OIII] luminosities given in Table 1. For the merger rates, I use the morphological classification in Table 2 and include all objects as disturbed that are listed to have tidal tails, I do not include objects listed as having a companion only to match the criteria outlines in Section \ref{S:f_merg}.

\subsection{Zakamska+19 (Z2, ZR)}

\citet{zakamska_host_2019} study 10 extremely red quasars (ERQs) and six Type 2 AGN at redshift $2<z<3$ using HST F814W and F160W. They do not list luminosities in any specific band and instead give bolometric luminosity ranges for their sample. This is used here. For the detected merger fraction, they use visual inspection. In the abstract, they cite the number of sources with close neighbours.  In the discussion, they additionally cite the fraction of sources showing tidal tails, to make these results more in line with our outlined merger fraction rules (see Section \ref{S:f_merg}), the latter is chosen to match the rules in Section \ref{S:f_merg}. This gives 0/6 for the Type 2 AGN and 1/10 for the ERQs.  

The bolometric luminosities for this sample are not explicitly given and their origins are unclear. The paper cites "a few times 10$^{46}$ [erg/s]" for the Type 2 sample and "typically 10$^{47}$, but in some cases reaching 10$^{48}$ [erg/s]" for the ERQs, both refer to the parent sample, rather than the sample itself. These rather broad ranges are used.

\begin{figure*}[h]
\begin{center}
\includegraphics[width=18cm]{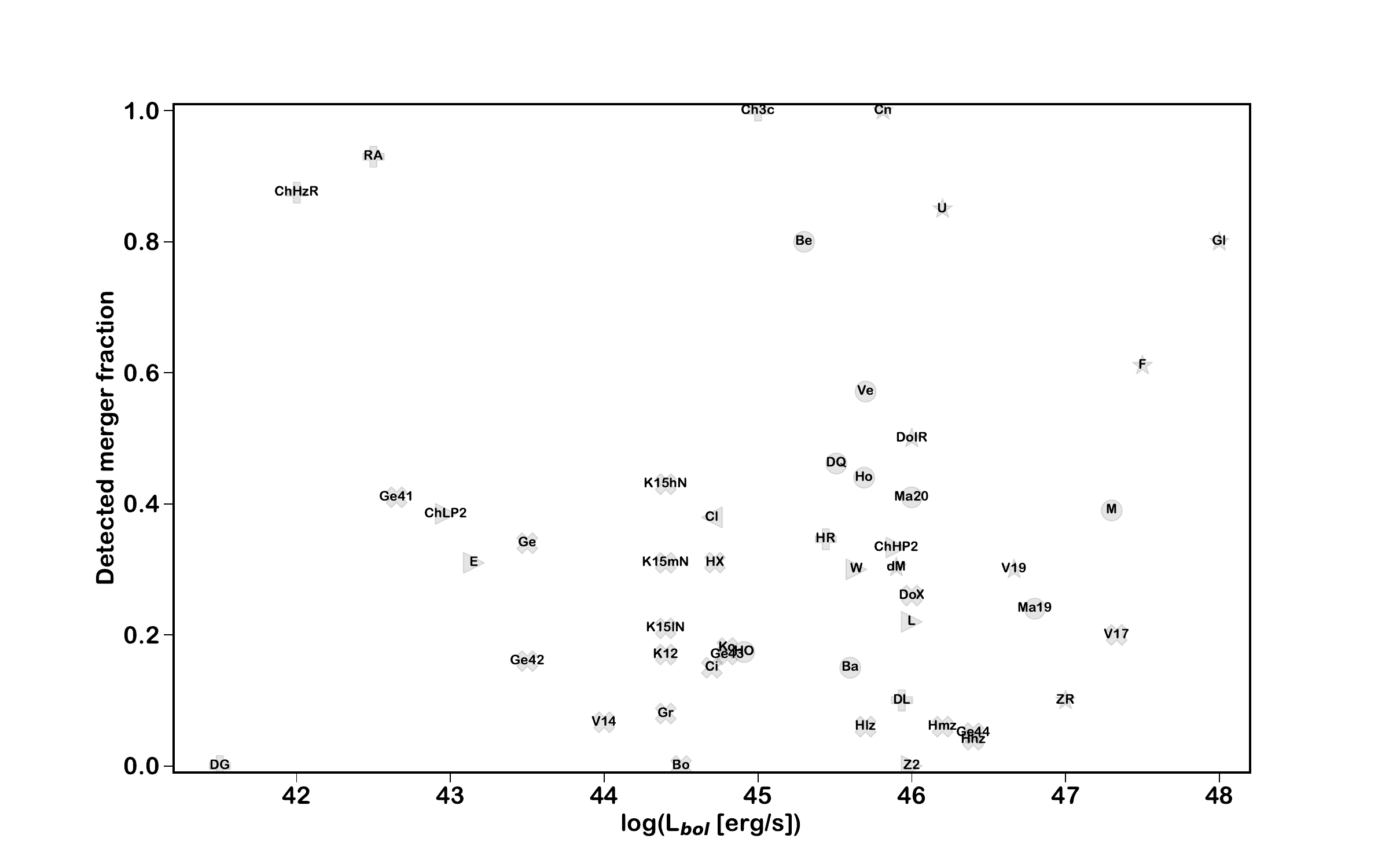}
\caption{Location of each study in Figure \ref{F:mainplot}, labels refer to the abbreviations used in Table \ref{maintable}.}
\label{F:labels}
\end{center}
\end{figure*}

\end{document}